\documentclass[pdflatex,sn-mathphys]{sn-jnl} 

\jyear{2022}%

\theoremstyle{thmstyleone}%
\theoremstyle{thmstyletwo}%

\theoremstyle{thmstylethree}%

\usepackage{siunitx} 
\usepackage{amsmath,amstext,color,tabu,mathtools}
\usepackage{gensymb}
\usepackage{natbib}
\usepackage{hyperref}
\usepackage{longtable}
\usepackage{graphicx}
\usepackage{tablefootnote}

\begin{document}

\makeatletter
\let\jnl@style=\rm
\def\ref@jnl#1{{\jnl@style#1}}

\makeatother
\let\astap=\aap
\let\apjlett=\apjl
\let\apjsupp=\apjs
\let\applopt=\ao

\newcommand{\actaa}{Acta Astron.}   
\newcommand{\araa}{Annu. Rev. Astron. Astrophys.}   
\newcommand{\areps}{Annu. Rev. Earth Planet. Sci.} 
\newcommand{\aar}{Astron. Astrophys. Rev.} 
\newcommand{\ab}{Astrobiol.}    
\newcommand{\aj}{Astron. J.}   
\newcommand{\ac}{Astron. Comput.} 
\newcommand{\apart}{Astropart. Phys.} 
\newcommand{\apj}{Astrophys. J.}   
\newcommand{\apjl}{Astrophys. J. Lett.}   
\newcommand{\apjs}{Astrophys. J. Suppl. Ser.}   
\newcommand{\ao}{Appl. Opt.}   
\newcommand{\apss}{Astrophys. Space Sci.}   
\newcommand{\aap}{Astron. Astrophys.}   
\newcommand{\aapr}{Astron. Astrophys. Rev.}   
\newcommand{\aaps}{Astron. Astrophys. Suppl.}   
\newcommand{\baas}{Bull. Am. Astron. Soc.}   
\newcommand{\caa}{Chin. Astron. Astrophys.}   
\newcommand{\cjaa}{Chin. J. Astron. Astrophys.}   
\newcommand{\cqg}{Class. Quantum Gravity}    
\newcommand{\epsl}{Earth Planet. Sci. Lett.}    
\newcommand{\frass}{Front. Astron. Space Sci.}    
\newcommand{\gal}{Galaxies}    
\newcommand{\gca}{Geochim. Cosmochim. Acta}   
\newcommand{\grl}{Geophys. Res. Lett.}   
\newcommand{\icarus}{Icarus}   
\newcommand{\jcap}{J. Cosmol. Astropart. Phys.}   
\newcommand{\jgr}{J. Geophys. Res.}   
\newcommand{\jgrp}{J. Geophys. Res.: Planets}    
\newcommand{\jqsrt}{J. Quant. Spectrosc. Radiat. Transf.} 
\newcommand{\lrca}{Living Rev. Comput. Astrophys.}    
\newcommand{\lrr}{Living Rev. Relativ.}    
\newcommand{\lrsp}{Living Rev. Sol. Phys.}    
\newcommand{\memsai}{Mem. Soc. Astron. Italiana}   
\newcommand{\mnras}{Mon. Not. R. Astron. Soc.}   
\newcommand{\nat}{Nature} 
\newcommand{\nastro}{Nat. Astron.} 
\newcommand{\ncomms}{Nat. Commun.} 
\newcommand{\nphys}{Nat. Phys.} 
\newcommand{\na}{New Astron.}   
\newcommand{\nar}{New Astron. Rev.}   
\newcommand{\physrep}{Phys. Rep.}   
\newcommand{\pra}{Phys. Rev. A}   
\newcommand{\prb}{Phys. Rev. B}   
\newcommand{\prc}{Phys. Rev. C}   
\newcommand{\prd}{Phys. Rev. D}   
\newcommand{\pre}{Phys. Rev. E}   
\newcommand{\prl}{Phys. Rev. Lett.}   
\newcommand{\psj}{Planet. Sci. J.}   
\newcommand{\planss}{Planet. Space Sci.}   
\newcommand{\pnas}{Proc. Natl Acad. Sci. USA}   
\newcommand{\procspie}{Proc. SPIE}   
\newcommand{\pasa}{Publ. Astron. Soc. Aust.}   
\newcommand{\pasj}{Publ. Astron. Soc. Jpn}   
\newcommand{\pasp}{Publ. Astron. Soc. Pac.}   
\newcommand{\raa}{Res. Astron. Astrophys.} 
\newcommand{\rmxaa}{Rev. Mexicana Astron. Astrofis.}   
\newcommand{\sci}{Science} 
\newcommand{\sciadv}{Sci. Adv.} 
\newcommand{\solphys}{Sol. Phys.}   
\newcommand{\sovast}{Soviet Astron.}   
\newcommand{\ssr}{Space Sci. Rev.}   
\newcommand{\uni}{Universe} 

\title[GRB\,211211A]{The case for a minute-long merger-driven gamma-ray burst from fast-cooling synchrotron emission}


\author*[1]{\fnm{Benjamin P.} \sur{Gompertz}}\email{b.gompertz@bham.ac.uk}
\author[2,3]{\fnm{Maria Edvige} \sur{Ravasio}}
\author[1]{\fnm{Matt} \sur{Nicholl}}
\author[2]{\fnm{Andrew J.} \sur{Levan}}
\author[4,5]{\fnm{Brian D.} \sur{Metzger}}
\author[1]{\fnm{Samantha R.} \sur{Oates}}
\author[6]{\fnm{Gavin P.} \sur{Lamb}}
\author[7]{\fnm{Wen-fai} \sur{Fong}}
\author[2,8]{\fnm{Daniele B.} \sur{Malesani}}
\author[7]{\fnm{Jillian C.} \sur{Rastinejad}}
\author[6]{\fnm{Nial R.} \sur{Tanvir}}
\author[6]{\fnm{Philip A.} \sur{Evans}}
\author[2,9]{\fnm{Peter G.} \sur{Jonker}}
\author[6]{\fnm{Kim L.} \sur{Page}}
\author[10]{\fnm{Asaf} \sur{Pe'er}}

\affil[1]{\orgdiv{Institute for Gravitational Wave Astronomy and School of Physics and Astronomy}, \orgname{University of Birmingham}, \orgaddress{\street{B15 2TT}, \country{UK}}}

\affil[2]{\orgdiv{Department of Astrophysics/IMAPP}, \orgname{Radboud University}, \orgaddress{\street{6525 AJ Nijmegen}, \country{The Netherlands}}}

\affil[3]{\orgdiv{INAF}, \orgname{Astronomical Observatory of Brera}, \orgaddress{\street{via E. Bianchi 46}, \postcode{23807}, \city{Merate (LC)}, \country{Italy}}}

\affil[4]{\orgdiv{Center for Computational Astrophysics}, \orgname{Flatiron Institute}, \orgaddress{\street{162 W. 5th Avenue}, \city{New York}, \postcode{10011}, \state{NY}, \country{USA}}}

\affil[5]{\orgdiv{Department of Physics and Columbia Astrophysics Laboratory}, \orgname{Columbia University}, \orgaddress{\city{New York}, \postcode{10027}, \state{NY}, \country{USA}}}

\affil[6]{\orgdiv{School of Physics and Astronomy}, \orgname{University of Leicester}, \orgaddress{\street{University Road}, \city{Leicester}, \postcode{LE1 7RH}, \country{UK}}}

\affil[7]{\orgdiv{Center for Interdisciplinary Exploration and Research in Astrophysics and Department of Physics and Astronomy}, \orgname{Northwestern University}, \orgaddress{\street{2145 Sheridan Road}, \city{Evanston}, \postcode{60208-3112}, \state{IL}, \country{USA}}}

\affil[8]{\orgdiv{Niels Bohr Institute}, \orgname{University of Copenhagen}, \orgaddress{\street{Jagtvej 128}, \postcode{2200}, \city{Copenhagen N}, \country{Denmark}}}

\affil[9]{\orgdiv{SRON}, \orgname{Netherlands Institute for Space Research}, \orgaddress{\street{Niels Bohrweg 4}, \postcode{NL-2333 CA}, \city{Leiden}, \country{the Netherlands}}}

\affil[10]{\orgname{Bar Ilan University}, \orgaddress{\city{Ramat Gan}, \country{Israel}}}

\abstract{For decades, gamma-ray bursts (GRBs) have been broadly divided into `long'- and `short'-duration bursts, lasting more or less than 2\,s, respectively. However, this dichotomy does not map perfectly to the two progenitor channels that are known to produce GRBs -- the merger of compact objects (merger-GRBs) or the collapse of massive stars (collapsar-GRBs). In particular, the merger-GRBs population may also include bursts with a short, hard $<$~2\,s spike and subsequent longer, softer extended emission (EE). The recent discovery of a kilonova -- the radioactive glow of heavy elements made in neutron star mergers -- in the 50~s-duration GRB 211211A further demonstrates that mergers can drive long, complex GRBs that mimic the collapsar population. Here we present a detailed temporal and spectral analysis of the high-energy emission of GRB 211211A. We demonstrate that the emission has a purely synchrotron origin, with both the peak and cooling frequencies moving through the $\gamma$-ray band down to the X-rays, and that the rapidly-evolving spectrum drives the EE signature at late times. The identification of such spectral evolution in a merger-GRB opens avenues for diagnostics of the progenitor type.
}

\maketitle

\begin{figure*}[t]
    \centering
    \includegraphics[width=\textwidth]{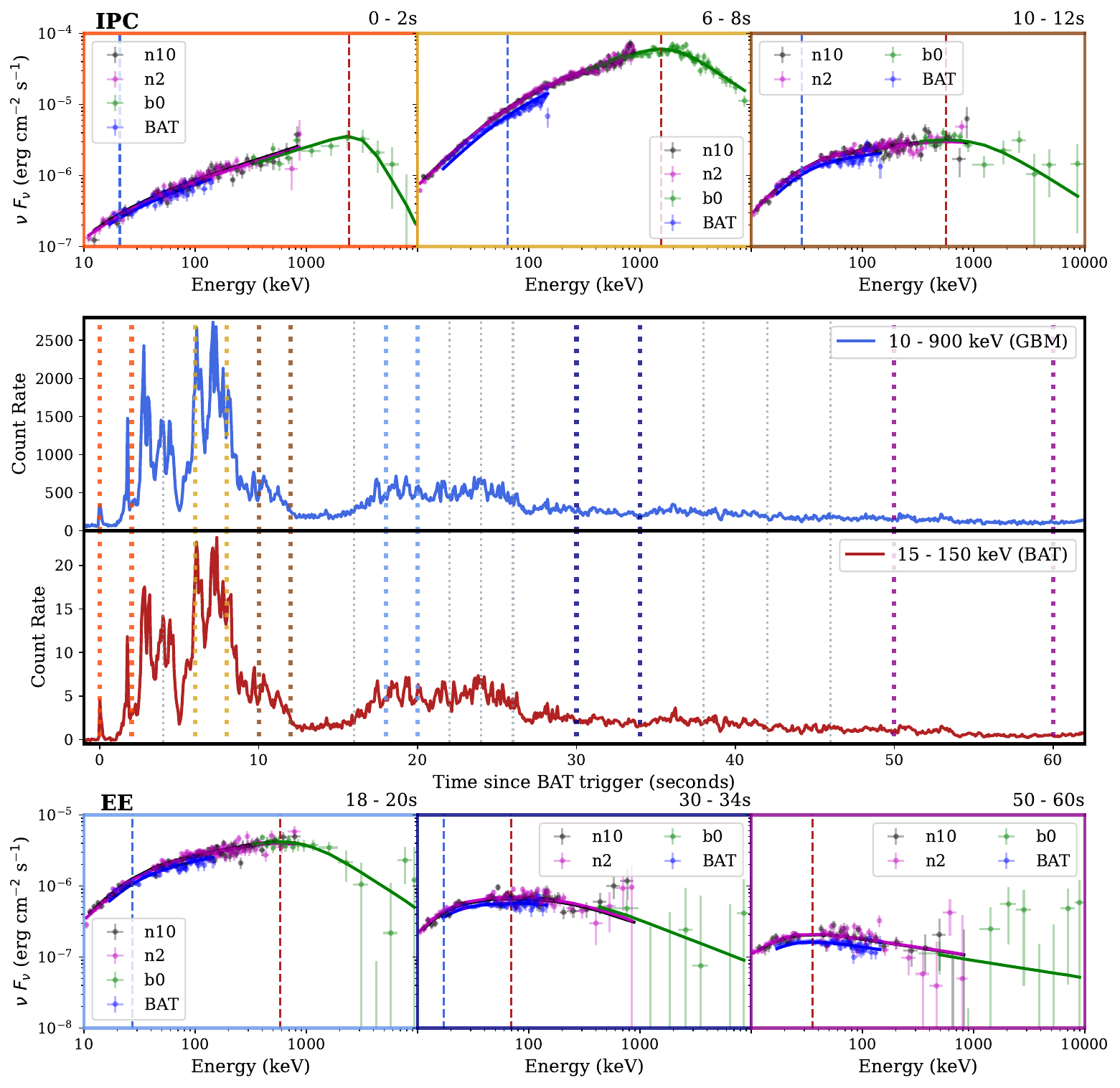}
    \caption{Spectral fits with the 2SBPL model from six representative epochs in the GBM and BAT light curves. Each spectral fit has n = 362 energy bins comprised of 3 GBM detectors (n10, n2 and b0) and BAT. Data are presented as mean values ± one standard deviation. The epochs bracketed by dotted lines in the light curve correspond to the spectra with frames of the same colour. In the spectral panels, the peak energy is shown by the dashed red line and the low energy break (where included in the fit) by the dashed blue line. The top row shows the highly variable nature of the IPC, while the EE phase along the bottom row exhibits a more gradual and coherent evolution.
    }
    \label{fig:spectra}
\end{figure*}

GRB 211211A was detected by \emph{Fermi}-GBM and \emph{Swift}-BAT at $t_0 =13$:09:59~UT on 2021 December 11. It was initially classified as a long burst due to its measured duration of $\approx 34.3$\,s (10\,keV -- 10\,MeV; \emph{Fermi} \cite{Mangan21}) and $51.4\pm0.8$\,s (15 -- 150\,keV; \emph{Swift} \cite{Stamatikos21}), in excess of the $2$\,s divide \cite{Kouveliotou93}. \emph{Swift} promptly slewed to point its X-ray Telescope (XRT; \cite{Burrows05}), which began settled observations of the field $79.2$\,s after the BAT trigger \cite{D'ai21}. X-ray observations showed bright emission ($0.3$ -- $10$\,keV flux $\approx 3\times10^{-8}$~erg~s$^{-1}$\,cm$^{-2}$) with an exponential decay lasting until $\sim 300$\,s after trigger, consistent with previous examples of EE (e.g. \cite{Norris02,Norris06,Norris10,Gompertz13}). The Ultra-Violet/Optical Telescope (UVOT; \cite{Roming05}) began pointed observations of GRB 211211A 92\,s after the BAT trigger \cite{Belles21}, and detected the optical counterpart. Late monitoring \cite{Rastinejad22} identified a significant near-IR excess which cannot be explained by radioactive Nickel decay but is remarkably similar to AT2017gfo, the kilonova discovered alongside GRB~170817, confirmed to be a binary neutron star (BNS) merger by gravitational-wave (GW) observations \cite{Abbott17_BNS}. The detection of a kilonova establishes that GRB 211211A originated from a compact merger (see Methods). Its association with a host galaxy at $z = 0.076$ \citep{Malesani21,Rastinejad22} makes it the closest confirmed compact object merger discovered without GWs, at a luminosity distance of just 350\,Mpc. Due to its proximity, GRB 211211A has the best-sampled high energy emission of any merger-driven GRB, and its surprisingly long duration affords us a remarkable opportunity to investigate the underpinning processes that enable compact object mergers to create long-duration GRBs. The burst was also detected at GeV frequencies \cite{Mei22,Zhang22}.

\begin{figure}[h]
    \centering
    \includegraphics[width=10cm]{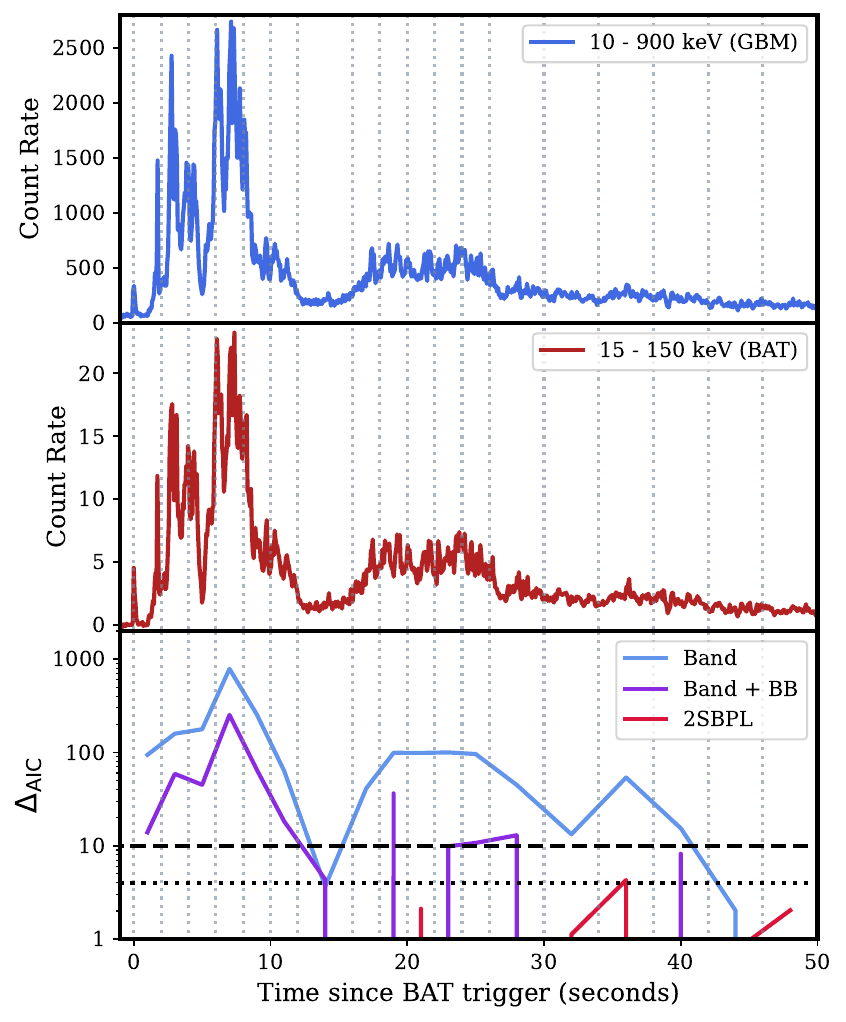}
    \caption{Relative support for the fitted models from Akaike's Information Criterion (AIC). Count rate light curves for GBM (top) and BAT (middle), showing the initial pulse complex ($t \lesssim 12$~s) and EE ($t \gtrsim 16$~s). The first pulse, on which both detectors triggered, is remarkably weak and is separated from the main emission episode by almost 2\,s. It may represent a `precursor' event before the merger. The lower panel shows the relative AIC ($\Delta_{\rm AIC}$) for the models described in Methods. The preferred model always has $\Delta_{\rm AIC} = 0$. Models with $\Delta_{\rm AIC} > 4$ (dotted line) are disfavoured, while those with $\Delta_{\rm AIC} > 10$ (dashed line) have no support from AIC. Spectral slices are indicated by the dotted grey lines.}
    \label{fig:redchi}
\end{figure}

We extract time-resolved spectra in narrow windows of 2--20\,s from the \emph{Swift} and \emph{Fermi} data and initially fit these with a range of spectral models using {\sc xspec v12.11.1} \citep{Arnaud96} (see Methods). Spectra from six representative epochs are shown in Figure~\ref{fig:spectra}, covering the spectrally hard Initial Pulse Complex (IPC) and softer EE. Although these episodes are distinct in the established nomenclature and are often discussed as such in this work, we do not discriminate between them in our fitting procedure, where we choose our spectral time bins based on signal-to-noise. The IPC and EE structure is reminiscent of GRB 060614, a $102$\,s long GRB at $z = 0.125$ that was strongly suspected to be a merger due to a lack of a supernova to deep limits \citep{Gehrels06}. Notably, the hard prompt emission is almost twice as long in GRB 211211A, further emphasising the diversity of merger-driven events. We find that the data are well described by a double smoothly broken power-law model (2SBPL; \cite{Ravasio18}, shown in Figure~\ref{fig:spectra}). A standard Band function \citep{Band93} with a single break is strongly disfavoured by Akaike’s Information Criterion \cite{Akaike74,Burnham04} (Figure~\ref{fig:redchi}). A Band model with an additional thermal component is also disfavoured with respect to 2SBPL, especially in the brightest pulses, and further requires a high intrinsic hydrogen column density, inconsistent with the burst's afterglow \cite{Rastinejad22} (see Methods). The model parameters derived from our full suite of fits are given in Table~\ref{tab:fit_results}. We focus the rest of our discussion on the 2SBPL model (though for epochs where one of the breaks lies outside the observed frequency range, this reduces to the Band model; see Methods).

\begin{figure}[h]
    \centering
    \includegraphics[width=10cm]{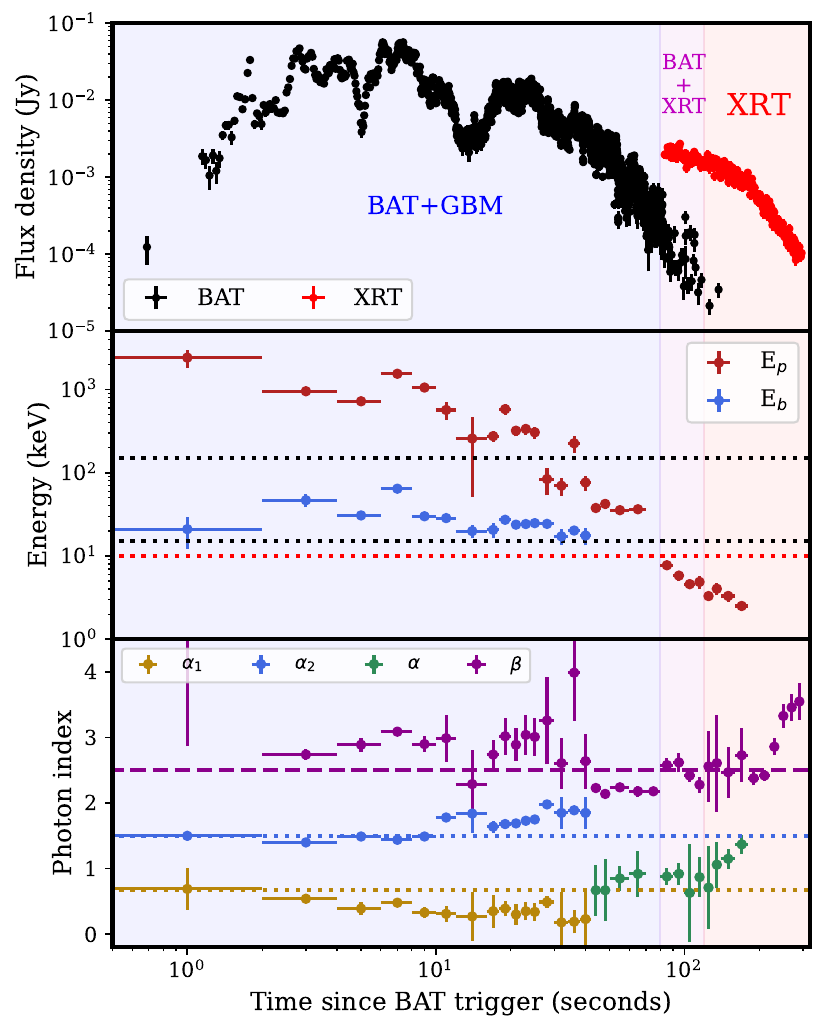}
    \caption{The evolution of the synchrotron spectrum from our time-resolved spectral fits. 2SBPL is fitted to the early data, but simpler models are used at late times where necessary. Shaded regions indicate the detectors available for fitting. All data are presented as mean values ± one standard deviation. \textbf{Top:} \emph{Swift}-BAT (n = 1892 data points) and XRT (n = 228 data points) 10\,keV flux density light curves of the prompt + EE phases. \textbf{Middle:} The evolution of E$_{p}$ (n = 28 measurement epochs) and E$_{b}$ (n = 16 measurement epochs). The BAT bandpass is marked by two horizontal black dotted lines. The horizontal red dotted line shows the lower limit of fitted GBM energies and the XRT upper bandpass limit. \textbf{Bottom:} Evolution of the photon indices derived from fitting (n$_{\alpha_1}$ = 16; n$_{\alpha_2}$ = 16; n$_{\alpha}$ = 12; n$_{\beta}$ = 35, where n is the number of measurement epochs). The expectation values of $\alpha_1$ and $\alpha_2$ in the fast-cooling regime and the nominal upper limit of $\beta$ are indicated by the dashed lines.}
    \label{fig:sync_evolution}
\end{figure}

The 2SBPL model parameterises a synchrotron spectrum, where relativistic charged particles radiate due to acceleration in magnetic fields. The spectrum is described by power-law segments connecting characteristic break frequencies \citep[see e.g.][]{Rybicki1979,Sari1998,Piran1999}. The three power-law segments have photon indices $\alpha_1$, $\alpha_2$ and $\beta$ (following the convention $\nu F_{\nu} \propto E^{2-\gamma}$, for a photon index $\gamma$), smoothly connected at two breaks, $E_b$ and $E_p$. These breaks map to the characteristic synchrotron frequency $\nu_m$, where most of the electrons radiate, or the cooling frequency $\nu_c$, above which electrons rapidly lose their energy. If $\nu_c < \nu_m$, all electrons cool down rapidly and the system is in the so-called `fast cooling' regime, while $\nu_m < \nu_c$ represents a `slow cooling' state (see Methods for more details).

\begin{figure}[h]
    \centering
    \includegraphics[width=10cm]{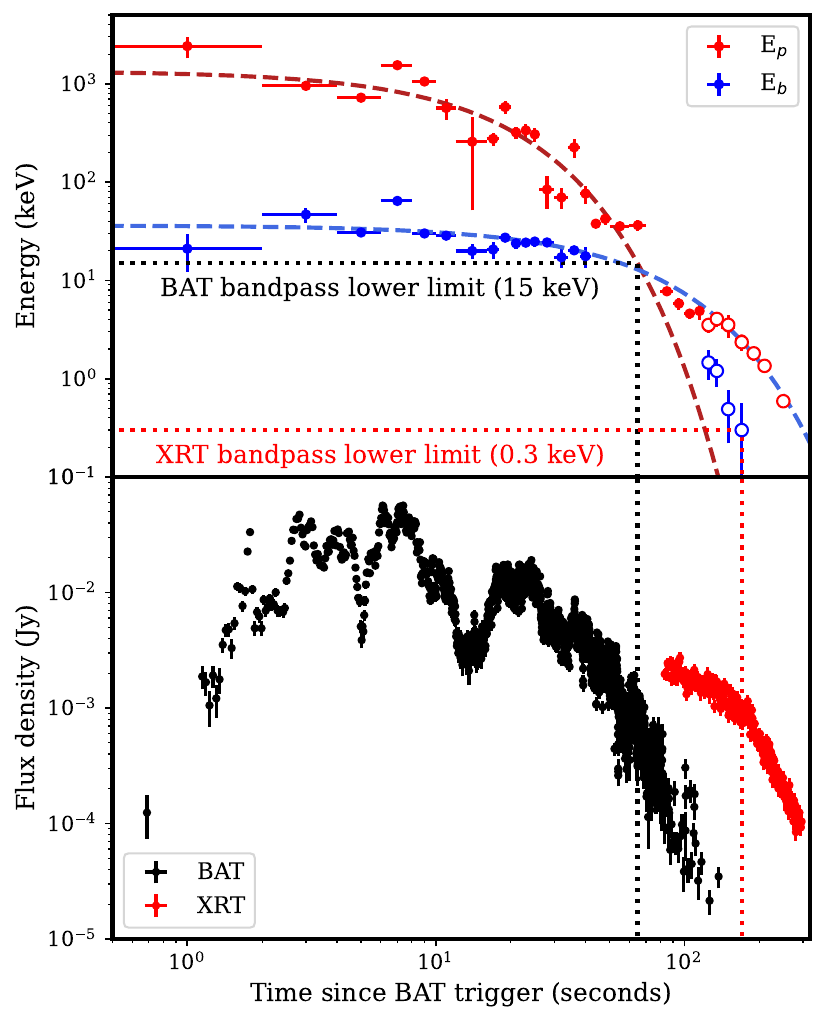}
    \caption{Evolution of the spectral breaks and their influence on the light curve. All data are presented as mean values ± one standard deviation. \textbf{Top:} The evolution of the spectral breaks $E_p$ (n = 24) and $E_b$ (n = 16) is well described by an exponential profile, where n is the number of spectral break measurements in each fit. Exponentials are fitted to the filled symbols of their respective colours. Open symbols are obtained by fixing the spectral slopes to their theoretical expectations, and are not fit. The affinity of the blue curve for the late red data therefore provides further support for the transition of the two breaks. \textbf{Bottom:} BAT (n = 1892) and XRT (n = 228) light curves, where n is the number of data points from each instrument. The rapid turnover in the light curves corresponds to the breaks exiting the bandpasses, implying that the duration of EE is sensitive to their evolution.}
    \label{fig:exponentials}
\end{figure}

The $\gamma$-ray and X-ray light curves of GRB 211211A and the evolution of the best-fit 2SBPL parameters are shown in Figure~\ref{fig:sync_evolution}. 
During the first 42\,s of evolution, we see a rapid decrease in $E_p$ from an initial $2411.3 \pm 602.0$\,keV to $76.3 \pm 15.9$\,keV. $E_b$ also shows a smooth but much slower decline from a maximum of $64.3 \pm 4.9$\,keV to $17.6 \pm 4.2$\,keV. We find mean indices of $\alpha_1 = 0.52 \pm 0.24$ and $\alpha_2 = 1.61 \pm 0.25$. These values are in excellent agreement with the expected values of 2/3 and 3/2 for synchrotron emission in the fast-cooling regime. Evidence for the presence of two breaks has been previously found in bright GRBs detected by both \emph{Fermi} \cite{Ravasio18,Ravasio19,Ronchi2020,Burgess2020,Toffano2021} and \emph{Swift} \cite{Oganesyan17,Oganesyan2018,Oganesyan2019}, but all were collapsar GRBs rather than merger-driven.

After 42\,s, $E_b$ is unresolved and we fit a single photon index, $\alpha$, below $E_p$. Initially, the fitted values of $\alpha$ are in excellent agreement with the expected value of $\alpha_1$ (Figure~\ref{fig:sync_evolution}, lower panel, green points), indicating that $E_b$ and $E_p$ are unresolved because they are too close together, rather than $E_b$ having left the bandpass. From $\sim$120\,s, the measured value of $\alpha$ begins to rise. We interpret the change as being due to the synchrotron breaks now moving away from one another (Methods), resulting in $\alpha_2$ dominating the measurement as $E_b$ approaches the lower limit of the observed bandpass and pushes $\alpha_1$ out of the spectrum.

Indeed, by fixing the photon indices to their expected values, we can once again resolve $E_b$ in our fits starting from 120s (Figure~\ref{fig:exponentials}, open symbols). The increasing separation between $E_b$ and $E_p$ and the coincident evolution of $\alpha$ from 2/3 towards 3/2 indicates that the more rapidly-evolving spectral break is now the lower energy of the two, in contrast with the first 42\,s of evolution. This apparent crossing of the two spectral breaks is consistent with the source having transitioned from a fast-cooling regime to slow cooling \citep{Rybicki1979,Sari1998}.

\begin{figure}[h]
    \centering
    \includegraphics[width=10cm]{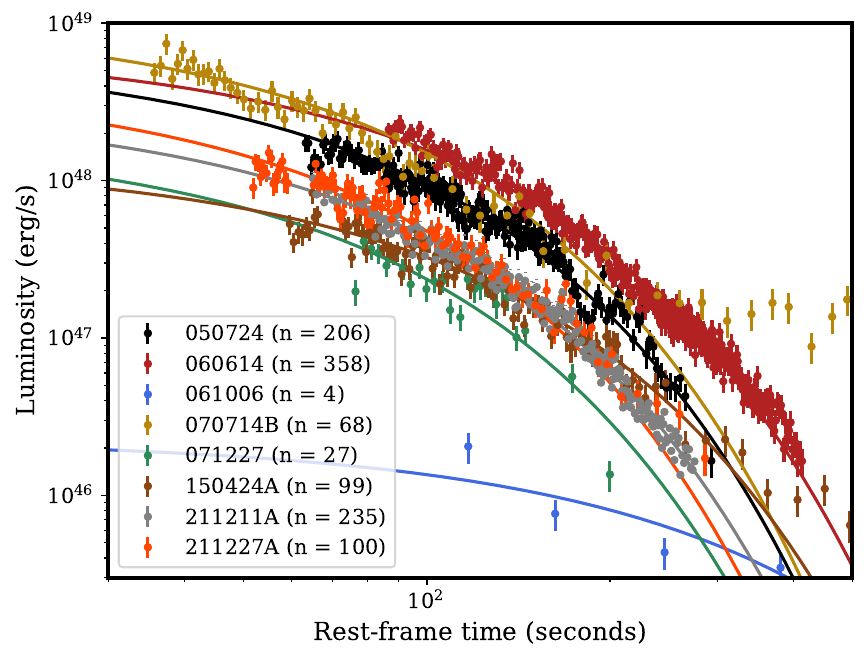}
    \caption{The X-ray light curves of EE GRBs. The sample is shown in Supplementary Table 1. Data are presented as mean values ± one standard deviation. The light curves have been fit with an exponential profile, and show remarkably consistent behaviour with the exception of the poorly-sampled GRB\,061006. Combined with the blueprint of spectral evolution found for GRB\,211211A in this work, this implies very uniform emission physics. The number of data points in each fit is given by n in the legend.}
    \label{fig:xrays_lcs}
\end{figure}

We further test this scenario by fitting an exponential profile of the form $E = Ne^{(-t/\tau)}$ to the evolution of both breaks (Figure~\ref{fig:exponentials}). The two profiles cross at $t_c = 68$\,s, and their crossing is further supported by the affinity of the $E_b$ fit to the $E_p$ measurements post-$t_c$. The evolution of the breaks has a marked effect on the EE light curve. The characteristic timescale for the evolution of $\nu_c$ ($\tau_{E_b} = 63.3 \pm 20.2$\,s) and the turnover of the X-rays ($\tau_X = 66.9 \pm 1.0$\,s) are close to the fast-to-slow cooling transition time ($t_c \approx 68$\,s) suggesting a connection between the cooling transition and the EE duration. Moreover, they are clearly driving the morphology of the late X-ray decline, which steepens when $\nu_m$ evolves below the band and ceases when $\nu_c$ follows (Figure~\ref{fig:exponentials}). The consistency of this X-ray feature across the EE class (e.g. \cite{Gompertz13}, Figure~\ref{fig:xrays_lcs}) hints that the shock physics may be remarkably uniform across the population of long-duration merger-GRBs.

At higher energies, the characteristic timescale of $\nu_m$ ($\tau_{E_p} = 14.4 \pm 1.14$\,s) is well matched to the duration of the IPC ($\sim 12$\,s). This likely relates to the duration of the jet, which energises the shock during the IPC, keeping $\nu_m$ consistently high. Although the duration of $\gamma$-ray EE is highly variable across the known sample (e.g. Supplementary Table 1), the evolution of $E_p$ through the BAT bandpass can explain the relative spectral softness of EE compared to the IPC (e.g. \cite{Norris10}). This framework also naturally leads to the population of GRBs that display the same exponential X-ray light curve as GRB 211211A and other EE GRBs (e.g. Figure~\ref{fig:xrays_lcs}), but lack EE in $\gamma$-rays \cite{Lu15,Gompertz20}. This can be achieved with a common evolution in $\nu_c$, while $\nu_m$ is found at lower energies below the BAT bandpass (after the IPC). Further, if $\nu_c$ is in fact \emph{below} the XRT bandpass in the majority of cases, one might recover a `canonical' short GRB lasting less than $2$\,s. However, this would imply a significant duration of `unseen' jet activity in these events.

Our analysis shows that the prompt emission spectrum of GRB 211211A is consistent with the so-called `marginally fast-cooling regime', with $\nu_c \lesssim \nu_m $ \citep{KumarMcMahon2008,Daigne11,Beniamini2013}. Detecting the cooling frequency $\nu_c$ at $\sim$ tens of keV implies that accelerated particles do not cool completely via synchrotron processes within a dynamical timescale \citep{Ghisellini2000}. By assuming that the electrons cool on a timescale of the order of the adiabatic one, we can derive self-consistent constraints on the magnetic field $B$ in the emitting region. For a typical emitting region radius $R$ and a bulk Lorentz factor of $\Gamma = 100$ (consistent with the afterglow; \cite{Rastinejad22}), our findings imply a magnetic field of the order of $B \sim 90 \, \nu_{c,60 \rm\,keV}^{-1/3} \, R_{13.5}^{-2/3} \, \Gamma_{2}$\,G, where $Q = Q_n \times 10^n$ (the magnetic field falls in the range $B\sim 30 - 200$ G for a range of $R\sim 10^{13}-10^{14}$\,cm). These values are consistent with those derived for marginally fast-cooling collapsar-GRBs \cite{Oganesyan2019, Oganesyan17, Oganesyan2018, Ravasio18, Ronchi2020}, but at odds with the expectations for a typical GRB emitting region ($B \sim 10^4 - 10^6$ G, e.g. \cite{Piran2005}).

\begin{figure}[h]
    \centering
    \includegraphics[width=10cm]{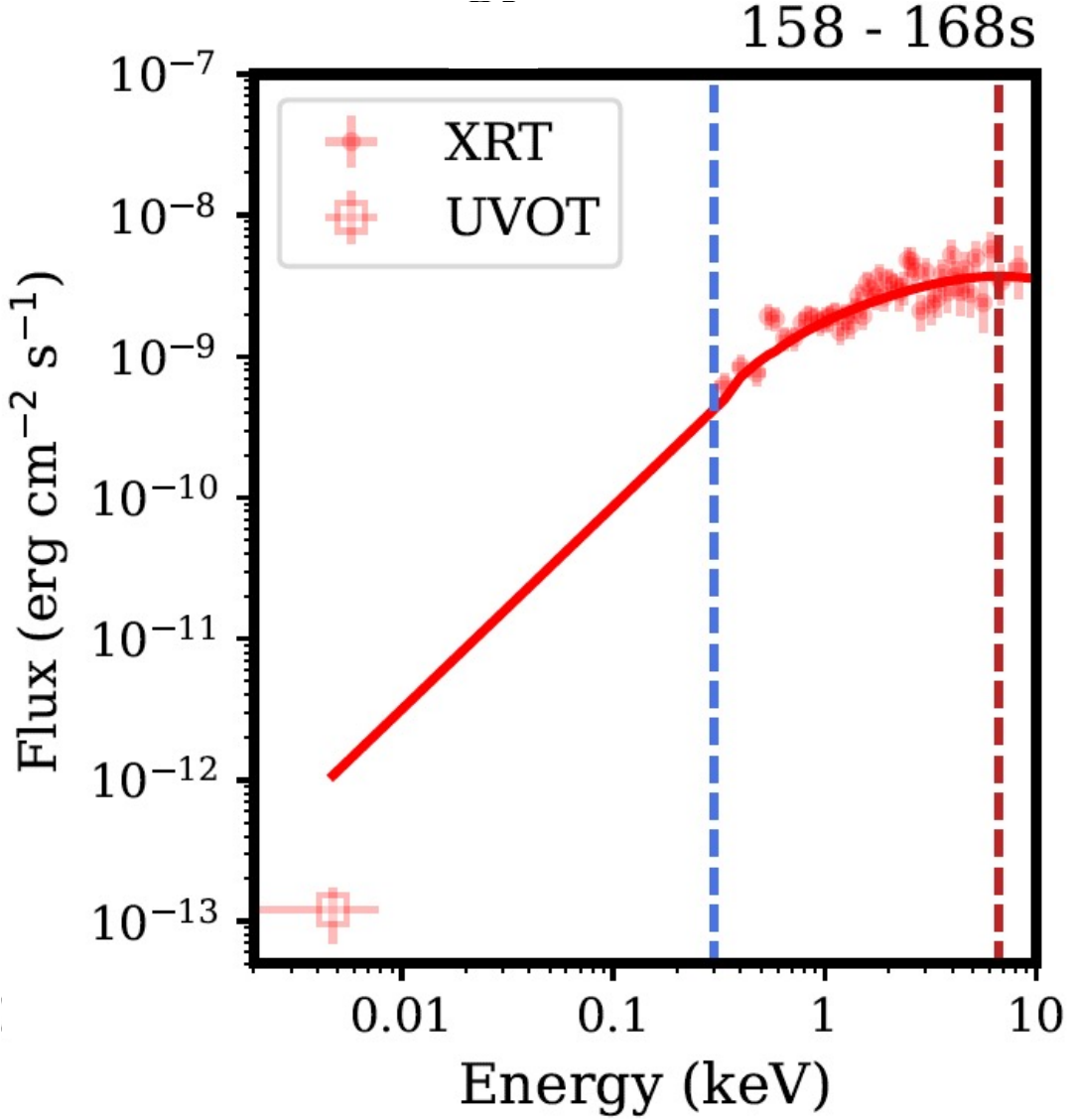}
    \caption{A joint fit of XRT (n = 483) and UVOT (n = 1) data centred around 163~s, where n is the number of energy bins contributed to the fit by each telescope. Data are presented as mean values ± one standard deviation. The red dashed line marks the position of $E_p$ in the fit, while $E_b$ is marked by the blue dashed line. The expected synchrotron photon index of 2/3 over-predicts the UV detection by an order of magnitude. This requires a spectral break or an additional source of absorption that cannot be explained by host galaxy extinction along the line of sight (see Methods).}
    \label{fig:xrt_uvot}
\end{figure}

Interestingly, we find that the UVOT data at $\sim$163\,s lie significantly below the extrapolation of the fitted synchrotron spectrum (see Methods, Figure~\ref{fig:xrt_uvot}). An additional spectral break is required to self-consistently model the spectral energy distribution (SED) from UV to X-rays. In the context of synchrotron emission, this can be interpreted as the synchrotron self-absorption frequency $\nu_{\rm ssa}$, below which the optical depth to synchrotron self-absorption is larger than unity. Following \cite{Shen2009} (but see also \cite{Sari1999,Granot2000,Granot2002,Beniamini2013}), we find that this implies a low $B$-field ($\sim 100$~G) and/or compact emitting region ($R < 10^{13}$~cm). However, a self-consistent solution that incorporates incomplete electron cooling and synchrotron self-absorption at UV frequencies is challenging, and may require alternative sources of absorption or modified synchrotron models (e.g. \cite{Ghisellini2020}) - particularly to avoid the bright synchrotron self-Compton emission predicted for such a system (see e.g. \cite{Ghisellini2020, KumarMcMahon2008,Beniamini2013}).

The identification of rapidly-evolving synchrotron emission as the driving force behind long-lived GRBs from mergers is an important step in understanding their genesis. Fundamentally, we do not know how a merger can produce $\sim$100s of emission in some cases, since this is well in excess of the expected $\lesssim 1$\,s accretion timescale of the post-merger torus (e.g. \cite{Lee09,Fernandez&Metzger13}). The two main explanations available in the literature are delayed fall-back accretion, or long-lived activity from the central engine.

Delayed fallback is typically invoked for a neutron star - black hole (NS-BH) merger, where a larger tidal tail of ejecta is expected (e.g. \cite{Foucart18}). In this case, EE might be related to the quantity and duration of the fall-back mass (e.g. \cite{Rosswog07,Desai19}). On the other hand, the characteristic $\sim 100$ s duration of the EE has no obvious explanation in the fall-back accretion scenario (however, see \cite{Kisaka15}). The kilonova associated with GRB 211211A is consistent with being a BNS merger \cite{Rastinejad22}, though an NS-BH cannot be completely ruled out. In the BNS scenario, EE may be powered by a temporarily or indefinitely stable magnetar remnant (e.g. \cite{Metzger08,Metzger11,Giacomazzo13,Gompertz14}) via a relativistic magnetized wind, potentially shaped by interaction with the merger ejecta into a collimated jet (e.g. \cite{Bucciantini12}). An abrupt change in jet properties is predicted to occur once the magnetar becomes optically-thin to neutrinos, on a timescale of $\sim 10-100$\,s \cite{Pons99}, which is suitable for EE. This transforms the composition of the magnetar outflow from an electron-ion plasma to an electron-positron outflow similar to that of a pulsar wind \cite{Metzger11}. Regardless of which, if either, model is the correct physical picture, establishing the observed evolution of $\nu_{\rm c}$ and $\nu_{\rm m}$ across the EE population represents the first step towards mapping these spectral features to the progenitor binary and post-merger remnant.

The brightness of GRB 211211A enables us to deconstruct both the initial prompt emission complex and the subsequent, softer extended emission. Our results suggest that the characteristics of EE in a merger-GRB with a long-lived engine depend mainly on where the synchrotron spectral breaks reside relative to the observational bandpasses. For the first ten seconds of the burst, $E_p$ and $E_b$ show little evolution. Therefore this period shows no spectral evolution and no spectral lag, a critical distinguishing feature of EE-GRBs \citep{Gehrels06}. After this time, $E_p$ softens rapidly, creating the observed softer EE component. Therefore, this evolution naturally explains both the lack of spectral lag in the early emission and the later softening seen in EE bursts. While only GRB 211211A has observations of the requisite quality to measure the motion of the spectral breaks directly, there is substantial similarity in spectral evolution and X-ray light curves in all EE bursts (see Methods). This implies that the same physical processes likely shape the entire class of events, and the 211211A template enables this possibility to be tested with a larger sample of merger- and collapsar-driven GRBs even at lower temporal resolution. If such behaviour can be established as unique to mergers through further studies, extended emission can be a smoking gun for the origin of a given GRB even in bursts at much larger distances or with much poorer localisations, e.g. in GW follow-up.

\setlength{\tabcolsep}{3pt}
\begin{table}[h]
\tiny
\centering
\begin{tabular}{ccccccccc}
\hline\hline
\textbf{2SBPL} \\
\hline
Epoch & $\alpha_1$ & $E_b$ & $\alpha_2$ & $E_p$ & $\beta$ & N$_{\rm H}$ & fit & dof \\
(s) & & (keV) & & (keV) & & (10$^{20}$\,cm$^{-2}$) & statistic & \\
\hline
1.0 $\pm$ 1.0 & 0.69 $\pm$ 0.32 & 21.0 $\pm$ 8.8 & 1.50 $\pm$ 0.06 & 2411.3 $\pm$ 602.0 & 4.74 $\pm$ 1.87 & --- & 315 & 353 \\
3.0 $\pm$ 1.0 & 0.54 $\pm$ 0.07 & 46.7 $\pm$ 8.2 & 1.40 $\pm$ 0.04 & 951.6 $\pm$ 47.4 & 2.74 $\pm$ 0.08 & --- & 480 & 353 \\
5.0 $\pm$ 1.0 & 0.39 $\pm$ 0.10 & 30.7 $\pm$ 3.3 & 1.49 $\pm$ 0.03 & 721.0 $\pm$ 47.5 & 2.89 $\pm$ 0.11 & --- & 401 & 353 \\
7.0 $\pm$ 1.0 & 0.48 $\pm$ 0.03 & 64.3 $\pm$ 4.9 & 1.44 $\pm$ 0.02 & 1543.0 $\pm$ 47.5 & 3.09 $\pm$ 0.07 & --- & 496 & 353 \\
9.0 $\pm$ 1.0 & 0.33 $\pm$ 0.09 & 30.0 $\pm$ 2.7 & 1.49 $\pm$ 0.02 & 1053.6 $\pm$ 72.0 & 2.90 $\pm$ 0.12 & --- & 378 & 353 \\
11.0 $\pm$ 1.0 & 0.31 $\pm$ 0.12 & 28.5 $\pm$ 2.4 & 1.78 $\pm$ 0.04 & 567.1 $\pm$ 135.7 & 2.99 $\pm$ 0.36 & --- & 387 & 353 \\
14.0 $\pm$ 2.0 & 0.27 $\pm$ 0.37 & 19.9 $\pm$ 3.6 & 1.84 $\pm$ 0.29 & 257.3 $\pm$ 205.7 & 2.29 $\pm$ 0.52 & --- & 378 & 353 \\
17.0 $\pm$ 1.0 & 0.35 $\pm$ 0.25 & 20.6 $\pm$ 4.2 & 1.64 $\pm$ 0.09 & 275.0 $\pm$ 40.9 & 2.74 $\pm$ 0.23 & --- & 391 & 353 \\
19.0 $\pm$ 1.0 & 0.39 $\pm$ 0.12 & 27.2 $\pm$ 2.8 & 1.68 $\pm$ 0.04 & 580.0 $\pm$ 86.5 & 3.02 $\pm$ 0.28 & --- & 389 & 353 \\
21.0 $\pm$ 1.0 & 0.30 $\pm$ 0.16 & 23.8 $\pm$ 2.9 & 1.69 $\pm$ 0.06 & 319.6 $\pm$ 45.8 & 2.89 $\pm$ 0.25 & --- & 366 & 353 \\
23.0 $\pm$ 1.0 & 0.35 $\pm$ 0.14 & 24.2 $\pm$ 2.6 & 1.73 $\pm$ 0.05 & 335.2 $\pm$ 51.5 & 3.04 $\pm$ 0.30 & --- & 363 & 353 \\
25.0 $\pm$ 1.0 & 0.34 $\pm$ 0.14 & 24.8 $\pm$ 2.7 & 1.75 $\pm$ 0.05 & 305.5 $\pm$ 47.9 & 3.01 $\pm$ 0.29 & --- & 353 & 353 \\
28.0 $\pm$ 2.0 & 0.49 $\pm$ 0.09 & 24.3 $\pm$ 0.9 & 1.98 $\pm$ 0.02 & 83.9 $\pm$ 30.3 & 3.26 $\pm$ 0.67 & --- & 393 & 353 \\
32.0 $\pm$ 2.0 & 0.18 $\pm$ 0.46 & 17.2 $\pm$ 3.7 & 1.85 $\pm$ 0.24 & 69.8 $\pm$ 16.9 & 2.61 $\pm$ 0.39 & --- & 341 & 353 \\
36.0 $\pm$ 2.0 & 0.19 $\pm$ 0.18 & 20.2 $\pm$ 1.6 & 1.89 $\pm$ 0.04 & 225.6 $\pm$ 50.9 & 3.99 $\pm$ 0.73 & --- & 381 & 353 \\
40.0 $\pm$ 2.0 & 0.23 $\pm$ 0.47 & 17.6 $\pm$ 4.2 & 1.85 $\pm$ 0.25 & 76.3 $\pm$ 15.9 & 2.64 $\pm$ 0.42 & --- & 331 & 353 \\
\hline\hline
\textbf{Band} \\
\hline
Epoch & $\alpha$ & & & $E_p$ & $\beta$ & N$_{\rm H}$ & fit & dof \\
(s) & & & & (keV) & & (10$^{20}$\,cm$^{-2}$) & statistic & \\
\hline
44.0 $\pm$ 2.0 & 0.67 $\pm$ 0.39 & --- & --- & 37.7 $\pm$ 3.06 & 2.23 $\pm$ 0.05 & --- & 334 & 355 \\
48.0 $\pm$ 2.0 & 0.67 $\pm$ 0.47 & --- & --- & 42.2 $\pm$ 5.61 & 2.14 $\pm$ 0.05 & --- & 337 & 355 \\
55.0 $\pm$ 5.0 & 0.85 $\pm$ 0.19 & --- & --- & 35.5 $\pm$ 2.29 & 2.24 $\pm$ 0.05 & --- & 384 & 355 \\
65.0 $\pm$ 5.0 & 0.92 $\pm$ 0.35 & --- & --- & 36.4 $\pm$ 4.32 & 2.18 $\pm$ 0.08 & --- & 317 & 355 \\
85.0 $\pm$ 5.0 & 0.88 $\pm$ 0.13 & --- & --- & 7.74 $\pm$ 0.81 & 2.58 $\pm$ 0.11 & $4.81 \pm 3.99$ & 373 & 453 \\
95.0 $\pm$ 5.0 & 0.92 $\pm$ 0.17 & --- & --- & 5.81 $\pm$ 0.83 & 2.62 $\pm$ 0.15 & $< 13.1$ & 468 & 530 \\
105.0 $\pm$ 5.0 & 0.63 $\pm$ 0.75 & --- & --- & 4.59 $\pm$ 0.54 & 2.42 $\pm$ 0.10 & $< 11.7$ & 442 & 445 \\
115.0 $\pm$ 5.0 & 0.87 $\pm$ 0.30 & --- & --- & 4.88 $\pm$ 0.89 & 2.28 $\pm$ 0.12 & $< 20.0$ & 468 & 616 \\
125.0 $\pm$ 5.0 & 0.71 $\pm$ 0.63 & --- & --- & 3.30 $\pm$ 0.37 & 2.56 $\pm$ 0.54 & $< 14.1$ & 310 & 408 \\
135.0 $\pm$ 5.0 & 1.06 $\pm$ 0.35 & --- & --- & 4.04 $\pm$ 0.67 & 2.61 $\pm$ 0.74 & $< 15.2$ & 375 & 567  \\
150.0 $\pm$ 10.0 & 1.15 $\pm$ 0.15 & --- & --- & 3.31 $\pm$ 0.49 & 2.47 $\pm$ 0.39 & $0$ & 372 & 427 \\
170.0 $\pm$ 10.0 & 1.37 $\pm$ 0.14 & --- & --- & 2.51 $\pm$ 0.30 & 2.73 $\pm$ 0.41 & $0$ & 348 & 478 \\
\hline\hline
\textbf{Power-law} \\
\hline
Epoch & & & & & $\beta$ & N$_{\rm H}$ & fit & dof \\
(s) & & & & & & (10$^{20}$\,cm$^{-2}$) & statistic & \\
\hline
75.0 $\pm$ 5.0 & --- & --- & --- & --- & 2.18 $\pm$ 0.04 & --- & 360 & 357 \\
190.0 $\pm$ 10.0 & --- & --- & --- & --- & 2.38 $\pm$ 0.11 & $9.52 \pm 2.20$ & 297 & 573 \\
210.0 $\pm$ 10.0 & --- & --- & --- & --- & 2.42 $\pm$ 0.08 & $8.33 \pm 1.51$ & 345 & 528 \\
230.0 $\pm$ 10.0 & --- & --- & --- & --- & 2.86 $\pm$ 0.13 & $8.46 \pm 2.03$ & 311 & 630 \\
250.0 $\pm$ 10.0 & --- & --- & --- & --- & 3.33 $\pm$ 0.18 & $11.6 \pm 2.64$ & 228 & 714 \\
270.0 $\pm$ 10.0 & --- & --- & --- & --- & 3.46 $\pm$ 0.20 & $14.0 \pm 3.05$ & 181 & 251 \\
290.0 $\pm$ 10.0 & --- & --- & --- & --- & 3.55 $\pm$ 0.28 & $13.4 \pm 4.08$ & 122 & 523 \\
\hline
Afterglow & --- & --- & --- & --- & 1.56 $\pm$ 0.07 & 1.19 $\pm$ 1.61 & 337 & 425 \\
\hline\hline
\end{tabular}
\caption{Parameters derived from our time-resolved spectral fitting. Fits are grouped by model type and so are not necessarily in chronological order. Data are presented as mean values ± one standard deviation except for the epoch column, which shows the central time and range of each spectral slice. $\alpha_1$, $\alpha_2$ and $\beta$ are spectral indices. $E_b$ and $E_p$ are break energies in keV. $N_H$ is the neutral hydrogen column number density along the line of sight in $10^{20}$\,cm$^{-2}$. The fit statistic represents the sum of the PGSTAT, C-stat and $\chi^2$ contributions, depending on which detectors were available in a given epoch.}
\label{tab:fit_results}
\end{table}




\section*{Methods}

\bmhead{Data products}

We retrieved GBM spectral data and their corresponding response matrix files (rsp2) from the online HEASARC archive \cite{Gruber14,vonKienlin14,vonKienlin20,Bhat16}. GBM count rate light curves are created using time-tagged event (TTE) data binned to a time resolution of 64~ms (Figure~\ref{fig:spectra}). No signal was detected by the Large Area Telescope (LAT) on board \emph{Fermi} during the prompt emission phase. The source incident angle was greater than 65$^{\circ}$ from the LAT boresight until $\sim$800\,s after the GBM trigger time.

We downloaded \emph{Swift} data from the UK \emph{Swift} Science Data Centre (UKSSDC; \cite{Evans07,Evans09}). Data from BAT are processed using the {\sc batgrbproduct} pipeline v2.48 from NASA's High Energy Astrophysics Software (HEAsoft v6.28; \cite{HEAsoft}). 15--150\,keV count-rate light curves are extracted using {\sc batbinevt} with 64~ms time resolution (Figure~\ref{fig:spectra}). Flux density light curves are taken directly from the UKSSDC Burst Analyser. We use the non-evolving data products binned to a signal-to-noise ratio of 4.

To assess spectral evolution, we extract time-resolved spectra in as narrow an interval as possible while retaining sufficient counts for the models to converge. We find that between 0 and 26\,s after the BAT trigger, binning the BAT and GBM data into 2\,s slices provides the best balance between cadence and signal-to-noise, except for a quiescent period at 12--16\,s, which sits between the prompt and EE phases (see Figure~\ref{fig:spectra}). This is combined into a 4\,s bin. Data between 26 and 50\,s are also grouped in 4\,s bins. Beyond 50\,s we use 10\,s bins. At 80\,s the GBM data end and the XRT data begin. We continue using 10\,s bins of BAT + XRT data up until the end of the BAT data at $\sim$120\,s after trigger. We then extract 10\,s bins of XRT data until 140\,s, where we widen our bins to 20\,s until the end of the EE phase at 300\,s.

\emph{Fermi}-GBM spectra were extracted using the public software {\sc gtburst}. We analysed the data of the two NaI detectors with the highest count rates and with the smallest viewing angle (n2, n10) and the most illuminated BGO detector (b0). We selected the energy channels in the range 10--900 keV for NaI detectors, excluding the channels in the range 30--40 keV (due to the iodine K–edge at 33.17 keV), and 0.3--40 MeV for the BGO detector. We modelled the background by manually selecting time intervals before and after the burst and fitting them with a polynomial function whose order is automatically found by {\sc gtburst}. 

After processing with {\sc batgrbproduct}, time-resolved BAT spectra are extracted using {\sc batbinevt}. These spectra have a systematic error vector applied via {\sc batphasyserr}, and {\sc batupdatephakw} is run to update the ray tracing keywords. We build a detector response matrix for each spectrum using {\sc batdrmgen}. XRT spectra are extracted using the UKSSDC webtool \citep{Evans07,Evans09}.

Fitting was performed using {\sc xspec v12.11.1} \citep{Arnaud96}. We include a free constant representing an effective area correction (EAC) to account for flux offsets due to the uncertain effective areas of the GBM detectors in flight. This is found to be a few per cent for intra-GBM calibration and up to $\sim 15$ per cent for GBM-BAT calibration.

\bmhead{Model comparison}

We fit our data with three models. The first and simplest is the Band function \citep{Band93}, which is the standard GRB continuum model and takes the form of two power-laws, with low-energy photon index $\alpha$ and a high-energy photon index $\beta$, smoothly connected at a $\nu F_{\nu}$ peak energy $E_p = E_c(2+\alpha)$:
\begin{multline}
\begin{split}
    N_E(E) &= A\bigg(\frac{E}{100  \textrm{ keV}}\bigg)^{\alpha} \exp\bigg(-\frac{E}{E_c}\bigg),&\\
    & &[(\alpha - \beta)E_c \geq E],\\
    &= A\bigg(\frac{(\alpha - \beta)E_c}{100 \textrm{keV}}\bigg)^{\alpha - \beta} \exp(\beta - \alpha)\bigg(\frac{E}{100 \textrm{keV}}\bigg)^{\beta},&\\
    & &[(\alpha - \beta)E_c \leq E]
\end{split}
\end{multline}
When there are no spectral breaks in the fitted bandpass, we simplify the Band function to a power-law fit corresponding to either $\alpha$ or $\beta$.

We also test the addition of a blackbody (BB) component, dividing the spectrum into thermal and non-thermal emission. The thermal component represents the emission from the photosphere of the GRB at typical radii of a few times $10^{11}$ cm from the central engine \citep{Pe'er15}. Non-thermal emission is produced further out ($\sim 10^{13}$ cm) either via interactions between expanding shells of material or through magnetic processes.

Finally, we fit a double smoothly broken power-law model \citep[2SBPL;][]{Ravasio18}. This model allows us to fit the spectra with three power-law segments (with photon indices $\alpha_1$, $\alpha_2$ and $\beta$) smoothly connected at two breaks (hereafter $E_b$ and $E_p$):
\begin{multline}
\begin{split}
    N_E(E) &= A E_b^{\alpha_1} \bigg[\bigg[\bigg(\frac{E}{E_b}\bigg)^{-\alpha_1n_1}+\bigg(\frac{E}{E_b}\bigg)^{-\alpha_2n_1}\bigg]^{\frac{n_2}{n_1}}\\
    &+ \bigg(\frac{E}{E_j}\bigg)^{-\beta n_2}\bigg[\bigg(\frac{E_j}{E_b}\bigg)^{-\alpha_1 n_1} + \bigg(\frac{E_j}{E_b}\bigg)^{-\alpha_2 n_1}\bigg]^{\frac{n_2}{n_1}}\bigg]^{-\frac{1}{n_2}},
\end{split}
\end{multline}
where
\begin{equation}
    E_j = E_p\bigg(-\frac{\alpha_2 + 2}{\beta + 2}\bigg)^{\frac{1}{(\beta-\alpha_2)n_2}}.
\end{equation}
This function represents the synchrotron spectral shape derived from a population of electrons assumed to be accelerated into a power-law distribution of energies, i.e. $N(\gamma_e) \propto \gamma^{-p}$ for $\gamma_{m} < \gamma_e < \gamma_{\rm max}$. The lower-energy $E_b$ and higher-energy $E_p$ represent either the cooling frequency ($\nu_c$, mostly emitted by electrons with Lorentz factor $\gamma_c$), or the characteristic synchrotron frequency ($\nu_m$, emitted by electrons at $\gamma_m$). 
The order of these breaks dictates the cooling regime: `slow cooling' ($\nu_m < \nu_c$), wherein only electrons with $\gamma_e > \gamma_c$ cool efficiently within a dynamical timescale, or `fast cooling' ($\nu_c < \nu_m$), in which all the electrons cool down rapidly to $\gamma_c < \gamma_m$. The two regimes have specific expectation values for the photon indices between the power-law segments.
For fast cooling ($\nu_c < \nu_m$), we expect $\alpha_1 = 2/3$ (for $\nu < \nu_c$), $\alpha_2 = 3/2$ (for $\nu_c < \nu < \nu_m$) and $\beta = p/2+1$ (for $\nu > \nu_m$), while for slow cooling ($\nu_m < \nu_c$) we expect $\alpha_1 = 2/3$ (for $\nu < \nu_m$), $\alpha_2 = (p+1)/2$ (for $\nu_m < \nu < \nu_c$), and $\beta = p/2+1$ (for $\nu > \nu_c$). For the typical parameters of the emitting region ($B \sim 10^4 - 10^6$~G, $\Gamma=100$), the particles are expected to efficiently radiate most of their energy on a timescale much smaller than the dynamical one and therefore the spectra are expected to be in fast-cooling regime \citep{Ghisellini2000}.

The 2SBPL function has been found to fit the spectra of the brightest \emph{Fermi} long GRBs significantly better than the standard Band-like single-break function \cite{Ravasio19}. The inclusion of the low-energy break (e.g. \cite{Oganesyan17, Oganesyan2018}) helps to solve the tension between the typically measured values of $\langle \alpha \rangle \approx 1$ in GRB spectra fitted with the Band function (e.g. \cite{Preece98}) and the expected values of $\alpha_1 = 2/3$ and $\alpha_2 = 3/2$ from synchrotron theory (e.g. \cite{Ghisellini2000}). This tension may also be lessened by considering inverse Compton effects in Klein-Nishina regime (e.g. \citep{Daigne11}). Recent simulations suggest that $\langle \alpha \rangle \approx 1$ may actually represent the mean value of these two unresolved spectral slopes \cite{Toffano2021}. Motivated by these results, we employ the 2SBPL function to test the consistency of GRB 211211A with a synchrotron-dominated spectrum.

Due to their different treatment of incident photons, the data from the three instruments used in our study must be fit with different fit statistics. For GBM, the background is modelled prior to extracting a spectrum, hence the resultant spectra are fit with PGSTAT, representing a Poissonian source over a Gaussian background. The mask weighting used in BAT means that its data must be fit with $\chi^2$ statistics. Finally, XRT employs straightforward photon counting, and hence is fit with Cash statistics \citep{Cash79}. As a result, different model fits cannot be reliably compared with $\chi^2$ statistic tools like the F-test. We therefore employ Akaike’s Information Criterion (AIC; \cite{Akaike74,Burnham04}) to compare models. We find that both the 2SBPL and Band+BB models provide a statistical improvement over the simpler Band function during the first $\sim 30$~s of data (Figure~\ref{fig:redchi}), with $\Delta_{\rm AIC} \gg 10$ indicating no support for the Band function \citep{Burnham04} (Supplementary Table 2). The 2SBPL model also provides a consistent improvement over Band+BB fits during the prompt emission phase (Supplementary Table 3).

In epochs that feature data taken with XRT, we include the effects of absorption from the neutral hydrogen column along the line-of-sight in our models. This is split into contributions from the Milky Way and the GRB host galaxy in the form {\sc [model]*tbabs*ztbabs} \citep{Wilms00}. Milky Way absorption is fixed to N$_{\rm H} = 1.76 \times 10^{20}$\,cm$^{-2}$ \citep{Willingale13}, with the contribution from the host galaxy at $z = 0.076$ left free to vary. We find that the Band+BB model requires the host N$_{\rm H}$ to be at least an order of magnitude higher than for fits with the 2SBPL model to achieve similar fit statistics.

To obtain the best measurement of the intrinsic N$_{\rm H}$, we perform fits to the time-integrated X-ray afterglow for the full set of XRT photon counting (PC) mode data (starting $\sim4000$\,s after trigger). The afterglow spectrum is well described (C-stat/dof = $337/422$) by a power-law model with a photon index of $\Gamma = 1.56 \pm 0.07$. It therefore provides a more reliable measurement of the absorption column than the rapidly evolving time-resolved spectra during EE. We find that N$_{\rm H} = (1.19 \pm 1.61) \times 10^{20}$\,cm$^{-2}$, consistent with zero and with a $3\sigma$ upper limit of N$_{\rm H} < 6.02 \times 10^{20}$\,cm$^{-2}$. By contrast, the brightest and best-sampled BAT+XRT bin at 80 -- 90\,s requires N$_{\rm H} = (2.55 \pm 0.40) \times 10^{21}$\,cm$^{-2}$ to achieve a good fit with the Band+BB model.

In short GRBs, the absorption column is expected to be constant over the lifetime of the source \citep[e.g.][]{Asquini19}. In rare cases, variations in N$_{\rm H}$ have been suggested for bright lGRBs \citep[e.g.][]{Campana21}, but the magnitude of the effect is at most a factor of two, and is sensitive to the choice of spectral model. In addition, an anomalously high neutral hydrogen column was also used to disfavour the thermal model in time-resolved spectral fitting of the EE GRB 060614 \citep{Mangano07}. We therefore rule out the Band+BB model firstly because our AIC test supports the 2SBPL model during the prompt phase, and secondly because it requires an anomalously high absorption column at X-ray frequencies. We focus the rest of our discussion on the 2SBPL model.

\bmhead{2SBPL fits}

We identify a general trend of variable spectral behaviour during the first $\sim 12$\,s (the IPC), which then gives way to a more coherent smooth evolution from the onset of EE ($\sim 20$\,s) onwards. We describe the evolution of the model parameters in various time windows here.\\
 
\noindent\textit{0 -- 80\,s: BAT + GBM.}

Between 0 and 80\,s our spectral coverage ranges from 10\,keV to 10\,MeV. We fit the full 2SBPL model for the first 42\,s of evolution. During this time, we see a rapid decrease in $E_p$ from an initial $2411.3 \pm 602.0$\,keV at 0--2\,s to $76.3 \pm 15.9$\,keV by 38--42\,s. $E_b$ also shows a smooth but much slower decline from an initial $21.0 \pm 8.8$\,keV to $17.6 \pm 4.2$\,keV over the same time range, though is first seen to rise to a peak energy of $64.3 \pm 4.9$\,keV at 6--8\,s after trigger.

In the same epoch, early values of $\alpha_1$ show excellent agreement with the expectation value of 2/3 before drifting downwards. The mean value of the low-energy photon index is $\overline{\alpha}_1 = 0.52 \pm 0.24$. The later discrepancy is still consistent with synchrotron expectations within $3\sigma$, and can be explained by $E_b$ falling close to the lower GBM bandpass, resulting in a poorly-sampled low-energy power-law segment. A similar issue has been found previously by \cite{Ravasio19} (see their figure B.1), who showed that all spectra with notably hard values of the low-energy photon index $\alpha_1$ also had a break energy of $E_b < 20$\,keV, suggesting an instrumental effect (the GBM threshold energy is $\sim$ 10\,keV). In addition, the smoothness parameter of the fitting function could influence the value of $\alpha_1$, especially at the edge of the detector bandpass where few energy channels are available to constrain it. We find that a sharper break results in $\alpha_1$ approaching the synchrotron-predicted value of 2/3. However, we retain a consistent smoothness across all of our fits to avoid biasing our results.

The slope $\alpha_2$ of the power-law between $E_b$ and $E_p$ is initially in excellent agreement with the expected value of $\alpha_2 = 3/2$ for fast cooling, but later evolves upwards towards a photon index of 2. This trend may be explained by uncertainties in distinguishing $\alpha_2$, $\beta$ and $E_p$ as the latter moves downwards though the band. In particular, as the source fades and the signal-to-noise ratio in the high energy GBM BGO detectors decreases, $\beta$ becomes poorly constrained. Our mean values are $\overline{\alpha}_2 = 1.61 \pm 0.25$ and $\overline{\beta} = 2.80 \pm 0.54$.

After 42\,s, the 2SBPL model is no longer statistically preferred over the Band function. For consistent fitting, we approximate the Band function by freezing $E_b$ below the bandpass, removing $\alpha_1$ and $E_b$ from the model. The fitted values of $\alpha$ are in excellent agreement with the expected value of $\alpha_1$ (Figure~\ref{fig:sync_evolution}, lower panel, green points), indicating that $E_b$ and $E_p$ are unresolved because they are too close together rather than $E_b$ having evolved out of the bandpass. We widen our time bins to 10\,s after 50\,s to improve the signal-to-noise of our spectra. Fitting the simpler Band model between 42 and 80\,s results in a much tighter constraint on the post-peak photon index, which now dominates the majority of the bandpass: $\overline{\beta} = 2.19 \pm 0.04$. This includes a single power-law fit in the 70--80\,s bin, indicating that both $E_p$ and $E_b$ are either below or very close to the 10\,keV lower bandpass limit of GBM.

We note that, assuming an isotropic-equivalent kinetic energy of the jet $E_{\rm k,iso} \sim 4 \times 10^{52}$ erg derived from the afterglow modelling reported in \cite{Rastinejad22} and the $\gamma$-ray isotropic-equivalent energy $E_{\rm iso} \sim 7.4 \times 10^{51}$ erg we observe in the energy range 10--1000 keV during the first 50 s of emission, the implied prompt $\gamma$-ray efficiency is $\eta = E_{\rm iso}/(E_{\rm iso}+E_{\rm k,iso}) \sim 16 \%$. This value is
consistent with estimates for other GRBs in the literature (e.g. \cite{Beniamini2016}), but requires a large contrast in the jet bulk Lorentz factor (e.g. \cite{Beloborodov2000}). We note however that recent works in the literature found a much smaller efficiency ($\eta \sim 10^{-3}$) when the fraction of electrons accelerated at the shock is included as a free parameter in the afterglow model (instead of the typical assumption that all electrons participate -- see e.g. \cite{Cunningham2020, Salafia2021}). These efficiencies are more compatible with the expectations of an internal shocks model with a mild contrast of bulk Lorentz factors \citep{Rees1994,Kumar1999, Spada2000}.

The magnetic field value implied by our findings ($B\sim 30$--200\,G for a range of $R\sim 10^{13}-10^{14}$\,cm) is consistent with the values derived in the cases of collapsar-GRBs showing marginally fast-cooling synchrotron spectra \citep{Oganesyan2019, Oganesyan17, Oganesyan2018, Ravasio18, Ronchi2020, Burgess2020}, but low compared to expectations for a typical GRB emitting region ($B \sim 10^4 - 10^6$ G, e.g. \cite{Piran2005}).
For such a low magnetic field in a small emitting region, synchrotron self-Compton (SSC) is expected to dominate the cooling rate \citep[see e.g.][]{Ghisellini2020, KumarMcMahon2008,Beniamini2013}. A larger emitting region ($R \sim 10^{15}-10^{16}$ cm) would reduce the SSC luminosity, but is more commonly associated with afterglow radiation and would be inconsistent with the rapid variability of the prompt light curve, unless the Lorentz factor is very high ($\Gamma \gtrsim 10^3$). One interesting solution to this problem has been recently proposed by \cite{Ghisellini2020}, ascribing the emission to proton-synchrotron in dominant adiabatic cooling. While this model offers an interesting alternative, it operates in a relatively small parameter space and its ability to explain all the observations is still under investigation \citep[see e.g.][]{Florou2021,Begue2021}.\\

\noindent\textit{80 -- 120\,s: BAT + XRT.}

After 80\,s there are too few GBM counts in 10\,s bins to obtain sufficient signal-to-noise. However, \emph{Swift}-XRT began pointed observations of GRB 211211A at $79.2$\,s after trigger \citep{D'ai21}, and we are able to fit combined BAT+XRT spectra between 0.3 and 150\,keV (excepting a 5\,keV inter-detector gap between 10 and 15\,keV). The extension of our spectral coverage down to $0.3$\,keV means that the Band function (approximated by 2SBPL with $E_b$ held below the XRT bandpass) is preferred over the single power-law model once more. The low energy index continues to be consistent with the expectations for $\alpha_1$, indicating that $E_b$ and $E_p$ remain unresolved from one another. $E_p$ continues to evolve smoothly downwards, from $E_p \sim 8$ keV to $\sim 5$ keV (see Figure~\ref{fig:sync_evolution}, Table~\ref{tab:fit_results}). The BAT data formally end at $t_0+122$\,s.\\

\noindent\textit{120 -- 300\,s: XRT and a possible transition to slow cooling.}

From $\sim$120\,s, the measured value of $\alpha$ begins to rise. We continue to measure a declining $E_p$ that has evolved to just a few keV and a consistent (but poorly constrained) $\beta$. We therefore interpret the change in $\alpha$ as being due to the synchrotron breaks now moving away from one another, resulting in an elongation of the 2SBPL $\alpha_2$ segment between them. Our measured $\alpha$ therefore evolves from being dominated by $\alpha_1$ to being a blend of $\alpha_1$ and $\alpha_2$ around the unresolved $E_b$. The power-law with index $\alpha_2$ then begins to dominate the measurement as it elongates, while $E_b$ approaches the lower bandpass limit and pushes $\alpha_1$ out of the spectrum.

We investigate this possibility by re-fitting the 2SBPL model with $\alpha_1$ and $\alpha_2$ fixed to their expected values (2/3 and 3/2, respectively), leaving the break energies and $\beta$ free. We find that between 50 -- 120\,s after trigger, the two breaks cannot be resolved from one another even with fixed photon indices. However, after 120\,s we measure a clear division again (Figure~\ref{fig:exponentials}, open symbols). The fixed indices mean that the spectral shape observed at this time is still consistent with the fast-cooling regime ($\nu_c < \nu_m$), but it is worth noting that the increasing separation between $E_b$ and $E_p$ and the coincident evolution of $\alpha$ from 2/3 towards 3/2 indicates that the more rapidly-evolving spectral break is now the lower energy of the two, in contrast with the first 42\,s of evolution. This apparent crossing of the two spectral breaks is consistent with the source having transitioned from a fast-cooling regime to slow cooling \citep{Rybicki1979,Sari1998} 
(see \cite{Burgess2020} for a possible slow-to-fast cooling transition).

If the spectrum is now slow cooling, $\alpha_2$ is no longer expected to have a value of 3/2, and instead should be $\alpha_2 = (p+1)/2$. We caution that slow cooling requires $p<3$. Given that $\beta = p/2 +1$, this is in contrast with the steep values of $\beta \gtrsim 2.5$ measured in the early spikes of emission, but is consistent with the values of $\beta$ measured at the implied time of transition ($\beta < 2.5$, Figure~\ref{fig:sync_evolution}). The transition from fast to slow cooling would therefore also require a change in the electron energy distribution power-law index $p$ over time.

For the epochs in which $E_b$ and $E_p$ are unresolved (which are covered by the Band fits in Table~\ref{tab:fit_results}) we find a mean $\overline{\beta} = 2.42 \pm 0.19$, which points to $p = 2.84 \pm 0.38$ and thus an expectation value of $\alpha_2 = 1.92 \pm 0.19$ in the slow-cooling regime. Unfortunately, beyond 180\,s our best fitting model reduces to a power-law while $\alpha_2$ is still evolving, meaning that we are unable to determine whether it supports fast or slow cooling. Our best-fitting 2SBPL model (which is not supported over the power-law fit by AIC) indicates $\alpha_2 = 1.68 \pm 0.12$, broadly consistent with either scenario. We note that at very late times $\beta$ evolves to extremely soft values ($\sim 3.5$). This is likely driven by an increase in free N$_{\rm H}$ (see Table~\ref{tab:fit_results}) and the decreasing signal-to-noise at higher energies in the bandpass.\\

\noindent\textit{UVOT: evidence for additional absorption?}

Regardless of the cooling regime, the spectral index below the break energy $E_b$ should be $\alpha_1 = 2/3$. A recent analysis of optical detections obtained during the prompt emission of 21 long GRBs revealed the spectrum between optical and $\gamma$-rays is well-modelled with synchrotron emission, i.e. optical data are consistent with the extrapolation of the expected 2/3 power-law index \cite{Oganesyan2019}.

The UVOT detection at $162.9 \pm 74.9$\,s allows us to investigate this possibility in GRB 211211A. We extract a 10\,s wide XRT spectrum centred at $162.9$\,s and perform a combined fit of the UVOT and XRT data. The UVOT data require the further addition of two {\sc zdust} components to the model to account for extinction in the GRB host galaxy and the Milky Way. The Milky Way extinction is fixed to $E(B-V) = 0.015$ \citep{Schlafly11}. Host extinction is fixed to $E(B-V) =0.18$, representing the $3\sigma$ upper limit that we derive from modelling the afterglow SED (see Methods: Afterglow SED).

To perform a joint fit of XRT and UVOT data, we use the 2SBPL model and fix $\alpha_1 = 2/3$, $\alpha_2 = 3/2$ and $E_b = 0.3$\,keV based on the results from our coincident spectral fit in Table~\ref{tab:fit_results}. $E_p$ was allowed to vary to obtain the best match to the X-ray data, since the fitted epoch is offset in time with respect to the template at $170 \pm 10$\,s. We find $E_p = 6.63 \pm 1.60$\,keV.

The extrapolation of $\alpha_1 = 2/3$ between the XRT and UVOT data over-predicts the UVOT flux by an order of magnitude (Figure~\ref{fig:xrt_uvot}). Larger values of extinction would be inconsistent with the afterglow SED observations; a free fit finds $E(B-V) = 0.60 \pm 0.06$, which is $\sim 7.5\sigma$ inconsistent with our model fit with the highest $E(B-V)$ (see Supplementary Table 4). An additional spectral break may therefore be required between the UVOT {\it white} filter and the XRT bandpass. One possibility for this break is the synchrotron self-absorption frequency ($\nu_{\rm ssa}$), below which photons are absorbed by the emitting electrons \citep{Rybicki1979}. This region of the spectrum has an expected photon index of $\alpha = -1$ (or $\nu^{-3/8}$ for an inhomogeneous distribution of electrons; \cite{Granot2000}), and hence can resolve the tension if it lies close to but above the UV.

The co-moving frame synchrotron self-absorption frequency can be computed by equating the co-moving synchrotron flux at $\nu_{ssa}^{\prime}$ to the flux from a blackbody in the Rayleigh–Jeans part of the spectrum. Following \cite{Shen2009}, we derived the distance $R$ of the emitting region from the central engine that is required to match the observations (i.e. that produce $6\times10^{-3}$\,keV $< \nu_{\rm ssa} < $ 0.3\,keV). The synchrotron self-absorption frequency $\nu_{\rm ssa}$ depends on the magnetic field of the emitting region $B$ and on the bulk Lorentz factor $\Gamma$ of the jet. In our computations, we required the magnetic field $B$ to be consistent with the one derived from the observed $\nu_{c} \sim 20 - 60$ keV during the early prompt emission (see Table \ref{tab:fit_results} and Figure~\ref{fig:sync_evolution}) and assumed the same bulk Lorentz factor as before: $\Gamma = 100$.

We find that a compact emitting region at $R \sim 10^{11-12}$~cm is required to place $\nu_{\rm ssa}$ between UV and X-rays. We note that the variability timescale implied by such a compact source is in contrast with the smooth emission $\sim 160$\,s after trigger. Moreover, such a small region would imply a high radiation density $U_{\rm rad}$, increasing the expected SSC luminosity and making it difficult to be optically thin
to Thomson scattering and pair production. A larger radius ($R \sim$ a few $10^{12-13}$~cm) can be accommodated while still having UVOT $< \nu_{ssa} <$ XRT by assuming $B \sim 10^4 - 10^6$ G, but this is inconsistent with the observation of incomplete cooling of the electrons. We therefore conclude it is unlikely that the UVOT flux is suppressed due to synchrotron self-absorption. Other causes of absorption (e.g. synchrotron photons absorbed by a cloud of not completely cooled electrons in the same line of sight, or pair-production with higher-energy photons) have to play a role in shaping the spectrum at low frequencies, but further, more quantitative investigations are needed to assess the feasibility of this process.

\bmhead{Afterglow SED}
We construct an optical-X-ray SED centred at $5.5{\rm ks}$. The SED was constructed using the methodology of \cite{sch10}, using data within a time range of 3 -- 10\,ks. To construct the pha files for the different optical filters at $5.5$\,ks, we first construct a single filter light curve by normalising the individual filter light curves together. Then using this single filter light curve we determine the temporal slope within the 3 -- 10\,ks interval. We then fit a power-law to the individual filter light curves within this same time interval, fixing the slope to the value determined from the single filter light curve. We use the derived normalisations to compute the count rate and count rate error at $5.5$\,ks, which was then applied to the relevant spectral file. For the XRT data, we take PC mode time-slice spectral files built using the UKSSDC webtool \cite{Evans09} within the 3 -- 10\,ks interval. The spectral files were normalised to correspond to the 3 -- 10\,ks flux of the afterglow at $5.5$\,ks. The flux, used to normalise a given spectrum, was determined by fitting a power-law to the temporal data within the SED selected time range. The best-fit decay index was used to compute the flux at $5.5$\,ks, in the same way as was done for the optical data. 

The SED was fit using {\sc xspec v12.11.1}, following the procedure outlined in \cite{schady07,sch10}. We tested two different models for the continuum: a power-law and a broken power-law, with the lower energy photon index fixed to be $\alpha = 2/3$, which corresponds to the expected spectral slope below the synchrotron peak frequency \citep{Sari1998}. In each of these models, we also included two dust and gas components to account for the Galaxy and host galaxy dust extinction and photoelectric absorption ({\sc phabs}, {\sc zphabs}, {\sc zdust}; we used {\sc zdust} for both the Galaxy and host galaxy dust components, but with the redshift set to zero for the Galaxy dust component and set to $z=0.076$ for the host components). The Galactic components were frozen to the reddening and column density values from \cite{schlegel} and \cite{kalberla}, respectively. For the host galaxy dust extinction, we tested the dependence of dust extinction on wavelength for three different scenarios: the Small and Large Magellanic clouds (SMC and LMC, respectively) and on the Milky Way (MW), finding no strong preference for any of these scenarios. In addition, we also account for the absorption due to the Lyman series in the $912-1215$\AA{} restframe wavelength range, with the {\sc XSPEC zigm} component. We set this component to use the prescription of \cite{mad95} to estimate the effective optical depth from the Lyman series as a function of wavelength and redshift and we set this component to also include attenuation due to photoelectric absorption. The results are provided in Supplementary Table 4.

\bmhead{Comparison to other GRBs}

The identification of a kilonova by \cite{Rastinejad22} means that despite its duration, GRB 211211A is a high-confidence merger event. While this discovery appears to answer the almost 20-year old question posed by the lack of supernovae in the nearby merger candidates GRBs 050724 and 060614 \citep{Barthelmy05b,Gehrels06}, 211211A is also not a typical EE GRB. In particular, the IPC is at least a factor of two longer than any previously identified example, which are generally consistent with the $t_{90} < 2$\,s hard spikes of the broader short GRB class (e.g. \cite{Perley09,Lien16}). We note that in this case both BAT and GBM triggered on an extremely faint spike almost 2\,s before the `main' episode of prompt emission (e.g. Figure~\ref{fig:spectra}). Such a weak pulse would likely not be detectable from much further away than GRB 211211A, and may represent a `precursor' event that artificially extended the IPC duration compared to other EE GRBs. Precursors like this have been suggested to arise from flares due to the resonant shattering of the crusts of pre-merger neutron stars \citep{Tsang12,Tsang13,Neill21}. Quasi-periodic oscillations have also been claimed in this feature \citep{Xiao22}.

We compare the broad properties of GRB 211211A with the sample of EE GRBs with redshifts from \cite{Gompertz20}, which we update to include EE GRB 211227A \citep{Lu22}. We note that \cite{Yang22} found GRB 211211A to be spectrally harder than previous EE GRBs during both the IPC and EE phases, though their Band function and cutoff power-law model fits do not capture all of the complex evolution presented here. GRB 211211A has both the shortest $t_{90}$ and highest $E_{\rm iso}$ of the known EE sample (Supplementary Table 1). 070714B has a shorter $t_{90}$ in the rest-frame and a comparable $E_{\rm iso}$, though accounting for redshift in the durations of GRBs has been shown to not be straightforward \cite{Littlejohns13}. Despite the diversity in $\gamma$-rays, the X-ray profile of EE is highly similar across all known events (Figure~\ref{fig:xrays_lcs}).

The sample of merger-driven GRBs investigated by \cite{Ravasio19} present a single power-law below the $\nu F_{\nu}$ peak, characterized by a hard photon index that is consistent with the expected 2/3 power-law below $\nu_c$ (if in fast cooling) or $\nu_m$ (if in slow cooling). In GRB 211211A, \emph{both} synchrotron breaks have been distinctively identified and tracked in detail, demonstrating that the cooling of electrons via synchrotron processes occurs in merger-driven GRBs. Conversely, the discovery of a power-law segment with photon index 3/2 between $\nu_c$ and $\nu_m$ in GRB 211211A is consistent with previous findings for collapsar GRBs, in addition to its long duration.

The $E_p$ measured in GRB 211211A is towards the higher end of the distribution found for the merger-driven GRBs in \cite{Ravasio19}, where the value of $E_p = 2411.3 \pm 602.0$ measured between 0 -- 2\,s would rank 2/11 (between $E_p = 2892.0^{+299.0}_{-205.0}$\,keV for GRB 120624A and $E_p = 1576.0^{+67.3}_{-67.1}$\,keV for GRB 090227B). This is also higher than the collapsar-driven sample. However, collapsar GRBs display a higher $E_b$ on average by a factor of several. In the extremely bright collapsar GRB 160625B, $E_p$ reached as high as $E_p = 6596.0^{+1220.0}_{-1400.0}$\,keV and $E_b = 196.7^{+31.2}_{-26.3}$\,keV at the peak of the main emission episode \citep{Ravasio19}. There is therefore no particular trend between our 2SBPL fit results for GRB 211211A and those given in the literature for collapsar- and merger-driven GRBs \citep{Ravasio18,Ravasio19}.

\bmhead{Evolution of the breaks}

The evolution of the spectral breaks is well-described by an exponential profile of the form $E = Ne^{(-t/\tau)}$ in both cases (Figure~\ref{fig:exponentials}). $E_p$ is best fit with $N_{E_p} = 1344 \pm 214$\,keV and $\tau_{E_p} = 14.4 \pm 1.14$\,s, while for $E_b$ we find $N_{E_b} = 36.2 \pm 4.39$\,keV and $\tau_{E_b} = 63.3 \pm 20.2$\,s. The two profiles cross at $t_c = 68$\,s, suggesting a transition from fast to slow cooling. Indeed, from $\sim$ 50\,s onwards (after the IPC and the peak of the EE luminosity), the spectral shape is consistent with a slow-cooling regime with $p \approx 2.2$ (Figure~\ref{fig:sync_evolution}). Measurements of $\beta > 2.5$ at early times ($\lesssim 50$\,s) are not consistent with this interpretation, and require that the index of the particles' energy distribution injected by the acceleration mechanism changes before the transition to slow cooling. Notably, $\beta$ is only seen to exceed $2.5$ during spikes in emission, suggesting that this may indeed be the case.

A change in cooling regimes has been claimed recently in \cite{Burgess2020}, where the authors found that single-pulse GRB spectra detected by \emph{Fermi}-GBM are well modelled with idealized synchrotron emission. The majority of the spectra are found to be in the slow-cooling regime, while a transition from slow to fast cooling has been found in some of them during the decay of the luminosity. The associated \emph{increase} in radiative efficiency of such a transition even as luminosity is seen to decrease further challenges the energetics in the jet. Conversely, the implied fast-to-slow cooling transition in GRB 211211A happens at late times, during the decaying phase of the luminosity. It is therefore more easily reconciled with the expected evolution of the jet.

Once the jet has ceased, the fading luminosity and downwards evolution of $E_p$ may be explained by a combination of adiabatic cooling \citep[cf.][]{Ronchini2021} and high-latitude effects. Within a structured jet, the energy dissipation that gives rise to the prompt emission may, in the wider regions where the bulk Lorentz factor is $\lesssim30$, occur below the photospheric radius, $R_p$. If this dissipation radius, $R_d$, is below the photosphere, photons will be coupled to the plasma until the optical depth reaches unity. Unless the condition for efficient thermalisation is met the resultant emission will have a lower luminosity and a spectral peak energy $\propto(R_p/R_d)^{-2/3}$ \citep{Lamb16}.

It is not clear how the exponential profile relates to EE in the BAT bandpass, though we note that Figure~\ref{fig:exponentials} suggests that $\tau_{E_p}$ may be under-estimated and perhaps includes the EE phase before the turnover. Indeed, the turnover of the BAT light curve coincides with the faster-evolving break (identified here as $\nu_m$) exiting the BAT bandpass. $t_{90}$, which measures the $\gamma$-ray duration exclusively, is highly variable across the EE sample (e.g. Supplementary Table 1), and this may point towards less uniform behaviour from $\nu_m$, corresponding to a greater diversity in IPC durations, which is indeed observed. However, $t_{90}$ is also subject to other effects like redshift or viewing angle, which would serve to obfuscate the implied correlation between the durations of the IPC and of EE at X-ray frequencies.

\bmhead{Evidence for a merger origin}

As shown by \cite{Rastinejad22}, the radio-to-X-ray observations following GRB 211211A can be readily explained by a kilonova with striking similarity to AT2017gfo \cite{Abbott17_BNS} superimposed over a GRB afterglow. The measured offset in projection from the putative host at $z = 0.076$ is $7.91 \pm 0.03$\,kpc, consistent with the median offset of $7.92$\,kpc for merger-GRBs \cite{Fong22}. The $i$-band upper limit at $17.6$ days after trigger excludes all known collapsar-GRB supernovae out to $z = 0.5$ \cite{Rastinejad22}. Measurements of the lag in arrival times between high- and low-energy photons are consistent with zero, and hence merger-GRBs \cite{Bernardini15}. The measured spectral lag is inconsistent with the established collapsar-GRB lag-luminosity relation \cite{Ukwatta10,Ukwatta12} unless the redshift is $z \gtrsim 1.5$ (corresponding to $L \geq 10^{53}$\,erg\,s$^{-1}$), but this is ruled out by the detection in the \emph{Swift}-UVOT UVM2 filter. A purely Ni$^{56}$-powered transient (a supernova or the merger of a neutron star with a white dwarf) is unable to evolve fast enough to match the GRB 211211A light curve.

Recently, \cite{Waxman22} proposed an alternative explanation for the infrared excess in GRB 211211A by invoking a dust destruction model in which a collapsar (originating from a massive star) was embedded in a dense molecular cloud that was subsequently heated by the UV radiation from the GRB. Their primary argument against the KN interpretation is the seemingly large K- to i-band flux ratio at $5.1$ days post-merger ($43 \pm 29$), but this is only in $1.2\sigma$ tension \cite{Rastinejad22} with the ratio seen in AT2017gfo (with significant systematic uncertainty) and comfortably within the range of radioactivity-powered models \cite{Kasen17}. In addition to the arguments above that strongly favour a merger-driven GRB, the dust destruction model requires that GRB 211211A is not associated with `Galaxy A' (whose light extends to the burst position) at $z = 0.076$ and instead originated from an unseen halo galaxy associated with Galaxy B at $z = 0.459$. This would be the faintest GRB host ever identified at $z < 3$ \cite{Lyman17} with a dust content that is atypical of very low mass galaxies \cite{Santini14,Calura17}. Further, we detect no residual absorption in either $N_H$ or $A_v$ over the Milky Way contribution in our study; the host contribution in both metrics is consistent with zero. We therefore conclude that the observed evolution and location of GRB 211211A strongly favours the merger interpretation.

\backmatter

\section*{Declarations}

\bmhead{Data Availability}

The majority of data generated or analysed during this study are included in this published article (and its supplementary information files). \emph{Swift} and \emph{Fermi} data may be downloaded from the UK \emph{Swift} Science Data Centre (\url{https://www.swift.ac.uk/}) and the online HEARSAC archive at \url{https://heasarc.gsfc.nasa.gov/W3Browse/fermi/fermigbrst.html}. Additional data is available upon reasonable request.

\bmhead{Code Availability}

The codes used in this publication are all publicly available. Fitting was performed in {\sc xspec} \cite{Arnaud96}, available from \url{https://heasarc.gsfc.nasa.gov/xanadu/xspec/}. \emph{Swift} tools are available from \url{https://heasarc.gsfc.nasa.gov/lheasoft/}, and \emph{Fermi} tools from \url{https://fermi.gsfc.nasa.gov/ssc/data/analysis/scitools/gtburst.html}. The 2SBPL model is published in \cite{Ravasio18}. Plots were created using {\sc matplotlib} \cite{Hunter07} in {\sc Python v3.9.7}.

\bmhead{Acknowledgments}
We thank the anonymous referees for careful consideration of our work and comments that improved the manuscript.

We thank Adam Goldstein for his insights into fitting GBM data. We thank Scott Barthelmy, David Palmer, Amy Lien and Donggeun Tak for discussions on the performance of BAT. We would like to thank Gabriele Ghisellini for fruitful discussions on emission physics.

The work makes use of data supplied by the UK \emph{Swift} Science Data Centre at the University of Leicester and the \emph{Neil Gehrels Swift Observatory}.

This research has made use of data obtained through the High Energy Astrophysics Science Archive Research Center Online Service provided by the NASA/Goddard Space Flight Center, and specifically this work has made use of public \emph{Fermi}-GBM data. 

B. Gompertz and M. Nicholl are supported by the European Research Council (ERC) under the European Union’s Horizon 2020 research and innovation programme (grant agreement No.~948381, MN). M. Nicholl acknowledges a Turing Fellowship.
A.J. Levan and D.B. Malesani are supported by the European Research Council (ERC) under the European Union’s Horizon 2020 research and innovation programme (grant agreement No.~725246, AJL). The Cosmic Dawn Center is funded by the Danish National Research Foundation under grant No{.} 140.
B.D. Metzger is supported in part by the National Science Foundation (grant Nos. Grants AST-2009255, AST-2002577).  The Flatiron Institute is supported by the Simons Foundation.
G. Lamb is supported by the UK Science Technology and Facilities Council grant, ST/S000453/1.
The Fong Group at Northwestern acknowledges support by the National Science Foundation under grant Nos. AST-1814782, AST-1909358 and CAREER grant No. AST-2047919. W.F. gratefully acknowledges support by the David and Lucile Packard Foundation. P.A. Evans and K.L. Page acknowledge funding from the UK Space Agency.  

\bmhead{Authors' Contributions}

B.P.G. extracted the \emph{Swift} data, performed the spectral analysis, provided the interpretation of the spectral evolution and the EE context, and wrote the text. M.E.R. extracted the \emph{Fermi} data, performed the spectral analysis, provided the interpretation on the emission physics, and co-wrote the text. M.N. and A.J.L. contributed vital insights into the direction of the study and co-wrote the text. B.D.M. provided theoretical interpretation, insights into the progenitor, and contributed to the text. S.R.O. performed the afterglow SED fits, reduced the UVOT data and contributed to the text. G.P.L. provided theoretical interpretation, self-consistency checks of the physics, and made contributions to the text. D.B.M. helped with the interpretation of the emission physics and in writing the text. W.F., J.C.R., N.R.T., P.G.J. and A.P. helped in the discussion and writing of the text. P.A.E. and K.L.P. provided insights into \emph{Swift} data reduction and handling, and commented on the text.

\bmhead{Conflict of interest}

We declare no conflicts of interests.

\bmhead{Supplementary Information}

Supplementary Information is available for this paper.





\clearpage


\end{document}


\renewcommand{\tablename}{Supplementary Table}

\section*{Supplementary Tables}

\begin{table}[h]
    \centering
    \begin{tabular}{cccc}
    \hline\hline
    GRB & $z$ & t$_{90}$ & E$_{\rm iso}$ \\
     & & (s) & (erg) \\
    \hline
    050724 & $0.257$ & $98.7 \pm 8.6$ & $1.71^{+0.13}_{-0.12} \times 10^{50}$ \\
    060614 & $0.125$ & $109.1 \pm 3.4$ & $7.65^{+0.36}_{-0.33} \times 10^{50}$ \\
    061006 & $0.438$ & $129.8 \pm 30.7$ & $6.74^{+0.41}_{-0.41} \times 10^{50}$ \\
    061210 & $0.41$ & $85.2 \pm 13.1$ & $4.22^{+0.42}_{-0.41} \times 10^{50}$ \\
    070714B & $0.923$ & $65.6 \pm 9.5$ & $1.16^{+0.09}_{-0.09} \times 10^{51}$ \\
    071227 & $0.381$ & $142.5 \pm 48.4$ & $1.96^{+0.30}_{-0.29} \times 10^{50}$ \\
    150424A & $0.3$ & $81.1 \pm 17.5$ & $6.25^{+0.65}_{-0.57} \times 10^{50}$ \\
    211211A & $0.076$ & $51.4 \pm 0.8$ & $2.06^{+0.03}_{-0.02} \times 10^{51}$ \\
    211227A & $0.228$ & $83.8 \pm 8.5$ & $1.14^{+0.03}_{-0.03} \times 10^{51}$ \\
    \hline\hline
    \end{tabular}
    \caption{Known EE GRBs from \cite{Gompertz13,Lien16,Gompertz20}. Values for $t_{90}$ are taken from \cite{Lien16,Lien21}. Here, $E_{\rm iso}$ is calculated in the 15 -- 150\,keV bandpass for consistency across all GRBs in the sample, and assumes a time-averaged spectrum. Data are presented as mean values ± one standard deviation.}
    \label{tab:EE}
\end{table}

\begin{table*}[ht]
\small
\centering
\begin{tabular}{ccccccc} 
\hline\hline
Epoch & $\alpha$ & $E_c$ & $\beta$ & fit statistic & dof & $\Delta_{\rm AIC}$ \\
(s) & & (keV) & & & & \\
\hline
$1.0 \pm 1.0$ & $1.41 \pm 0.02$ & $3561.9 \pm 1034.6$ & $< 3.4 \times 10^5$ & $413$ & $355$ & $94$ \\
$3.0 \pm 1.0$ & $0.92 \pm 0.01$ & $624.4 \pm 25.6$ & $2.35 \pm 0.04$ & $642$ & $355$ & $159$ \\
$5.0 \pm 1.0$ & $1.02 \pm 0.01$ & $453.5 \pm 23.2$ & $2.36 \pm 0.05$ & $582$ & $355$ & $177$ \\
$7.0 \pm 1.0$ & $0.86 \pm 0.01$ & $716.9 \pm 20.3$ & $2.22 \pm 0.02$ & $1286$ & $355$ & $786$ \\
$9.0 \pm 1.0$ & $1.02 \pm 0.01$ & $471.9 \pm 26.8$ & $2.15 \pm 0.03$ & $637$ & $355$ & $255$ \\
$11.0 \pm 1.0$ & $0.42 \pm 0.09$ & $50.0 \pm 6.04$ & $1.85 \pm 0.02$ & $455$ & $355$ & $63$ \\
$14.0 \pm 2.0$ & $0.16 \pm 0.23$ & $23.3 \pm 4.62$ & $1.92 \pm 0.02$ & $386$ & $355$ & $4$ \\
$17.0 \pm 1.0$ & $1.14 \pm 0.03$ & $210.1 \pm 21.7$ & $2.26 \pm 0.08$ & $435$ & $355$ & $41$ \\
$19.0 \pm 1.0$ & $1.00 \pm 0.03$ & $184.7 \pm 16.2$ & $2.05 \pm 0.04$ & $492$ & $355$ & $99$ \\
$21.0 \pm 1.0$ & $1.07 \pm 0.03$ & $180.9 \pm 15.7$ & $2.18 \pm 0.06$ & $466$ & $355$ & $99$ \\
$23.0 \pm 1.0$ & $1.00 \pm 0.04$ & $133.1 \pm 12.7$ & $2.05 \pm 0.04$ & $467$ & $355$ & $100$ \\
$25.0 \pm 1.0$ & $0.80 \pm 0.05$ & $83.7 \pm 8.32$ & $1.97 \pm 0.03$ & $453$ & $355$ & $96$ \\
$28.0 \pm 2.0$ & $0.76 \pm 0.09$ & $50.0 \pm 5.96$ & $2.09 \pm 0.03$ & $442$ & $355$ & $45$ \\
$32.0 \pm 2.0$ & $0.72 \pm 0.14$ & $37.2 \pm 6.17$ & $2.09 \pm 0.03$ & $357$ & $355$ & $13$ \\
$36.0 \pm 2.0$ & $0.67 \pm 0.13$ & $42.0 \pm 6.79$ & $1.96 \pm 0.02$ & $434$ & $355$ & $54$ \\
$40.0 \pm 2.0$ & $0.66 \pm 0.14$ & $36.0 \pm 5.92$ & $2.07 \pm 0.03$ & $351$ & $355$ & $15$ \\
$44.0 \pm 2.0$ & $0.58 \pm 0.23$ & $25.0 \pm 5.58$ & $2.16 \pm 0.03$ & $339$ & $355$ & $2$ \\
$48.0 \pm 2.0$ & $0.55 \pm 0.28$ & $23.8 \pm 6.49$ & $2.07 \pm 0.03$ & $339$ & $355$ & $0$ \\
\hline\hline
\end{tabular} 
\caption{Best spectral fits obtained with the Band function during the IPC and EE phases. The Band function features a low and high energy photon index $\alpha$ and $\beta$, respectively. $E_c$ is the characteristic energy, related to the peak by $E_c = E_p / (2 + \alpha)$. The fit statistic is the sum of the PGSTAT and $\chi^2$ contributions from the GBM and BAT spectral fits. The $\Delta_{\rm AIC}$ column gives the AIC relative to the best fitting model, which is 2SBPL in the majority of cases (see Supplementary Table \ref{tab:bandbbfits}). $\Delta_{\rm AIC} > 4$ indicates the Band model is disfavoured, with $\Delta_{\rm AIC} > 10$ indicating no support at all. Data are presented as mean values ± one standard deviation.}
\label{tab:bandfits}
\end{table*}

\setlength{\tabcolsep}{3pt}
\begin{table*}[ht]
\tiny
\centering
\begin{tabular}{ccccccccc} 
\hline\hline
Epoch & $\alpha$ & $E_c$ & $\beta$ & kT & R$_{bb}$ & fit & dof & $\Delta_{\rm AIC}$ \\
(s) & & (keV) & & (keV) & ($10^{10}$\,cm) & statistic & & \\
\hline
$1.0 \pm 1.0$ & $1.30 \pm 0.03$ & $3019.9 \pm 808.8$ & $< 2.5 \times 10^5$ & $14.6 \pm 2.4$ & $0.80 \pm 0.30$ & $329$ & $353$ & $14$ \\
$3.0 \pm 1.0$ & $0.95 \pm 0.01$ & $828.9 \pm 42.1$ & $2.54 \pm 0.06$ & $23.1 \pm 1.4$ & $0.38 \pm 0.04$ & $539$ & $353$ & $59$ \\
$5.0 \pm 1.0$ & $1.00 \pm 0.02$ & $598.6 \pm 38.1$ & $2.59 \pm 0.09$ & $15.2 \pm 0.7$ & $0.88 \pm 0.08$ & $446$ & $353$ & $45$ \\
$7.0 \pm 1.0$ & $0.85 \pm 0.01$ & $1005.7 \pm 29.4$ & $2.54 \pm 0.04$ & $22.8 \pm 0.6$ & $0.49 \pm 0.03$ & $748$ & $353$ & $252$ \\
$9.0 \pm 1.0$ & $1.07 \pm 0.02$ & $940.0 \pm 63.8$ & $2.57 \pm 0.09$ & $17.8 \pm 0.7$ & $0.76 \pm 0.06$ & $444$ & $353$ & $65$ \\
$11.0 \pm 1.0$ & $0.83 \pm 0.08$ & $181.4 \pm 27.2$ & $2.13 \pm 0.06$ & $10.5 \pm 0.5$ & $1.86 \pm 0.15$ & $405$ & $353$ & $18$ \\
$14.0 \pm 2.0$ & $1.02 \pm 0.17$ & $130.8 \pm 47.9$ & $2.03 \pm 0.08$ & $8.0 \pm 0.7$ & $2.86 \pm 0.36$ & $382$ & $353$ & $4$ \\
$17.0 \pm 1.0$ & $0.95 \pm 0.08$ & $192.6 \pm 27.2$ & $2.35 \pm 0.10$ & $8.7 \pm 0.6$ & $2.22 \pm 0.28$ & $390$ & $353$ & $0$ \\
$19.0 \pm 1.0$ & $0.93 \pm 0.06$ & $241.4 \pm 28.1$ & $2.22 \pm 0.06$ & $10.3 \pm 0.5$ & $1.71 \pm 0.18$ & $425$ & $353$ & $36$ \\
$21.0 \pm 1.0$ & $0.99 \pm 0.06$ & $244.1 \pm 28.1$ & $2.46 \pm 0.11$ & $10.4 \pm 0.5$ & $1.92 \pm 0.17$ & $363$ & $353$ & $0$ \\
$23.0 \pm 1.0$ & $1.01 \pm 0.05$ & $236.6 \pm 26.3$ & $2.48 \pm 0.11$ & $9.9 \pm 0.4$ & $2.03 \pm 0.18$ & $373$ & $353$ & $10$ \\
$25.0 \pm 1.0$ & $1.00 \pm 0.06$ & $244.7 \pm 25.6$ & $2.47 \pm 0.10$ & $10.3 \pm 0.5$ & $1.92 \pm 0.17$ & $364$ & $353$ & $11$ \\
$28.0 \pm 2.0$ & $1.16 \pm 0.06$ & $178.3 \pm 22.2$ & $2.71 \pm 0.21$ & $9.1 \pm 0.4$ & $2.59 \pm 0.21$ & $406$ & $353$ & $13$ \\
$32.0 \pm 2.0$ & $1.14 \pm 0.11$ & $125.5 \pm 26.7$ & $2.39 \pm 0.14$ & $7.5 \pm 0.5$ & $3.30 \pm 0.34$ & $340$ & $353$ & $0$ \\
$36.0 \pm 2.0$ & $1.09 \pm 0.07$ & $195.2 \pm 25.9$ & $2.99 \pm 0.35$ & $8.2 \pm 0.3$ & $3.20 \pm 0.23$ & $377$ & $353$ & $0$ \\
$40.0 \pm 2.0$ & $1.33 \pm 0.07$ & $213.4 \pm 35.2$ & $3.43 \pm 1.14$ & $8.8 \pm 0.5$ & $2.84 \pm 0.30$ & $340$ & $353$ & $8$ \\
$44.0 \pm 2.0$ & $1.36 \pm 0.10$ & $161.3 \pm 31.6$ & $< 1064.4$ & $7.0 \pm 0.4$ & $4.73 \pm 0.53$ & $333$ & $353$ & $0$ \\
$48.0 \pm 2.0$ & $0.89 \pm 0.44$ & $61.3 \pm 38.2$ & $2.14 \pm 0.07$ & $5.5 \pm 0.9$ & $4.34 \pm 0.71$ & $335$ & $353$ & $<1$ \\
\hline\hline
\end{tabular} 
\caption{Best spectral fits obtained with the Band + blackbody model during the IPC and EE phases. $E_c$ is the characteristic energy, related to the peak by $E_c = E_p / (2 + \alpha)$. The fit statistic is the sum of the PGSTAT and $\chi^2$ contributions from the GBM and BAT spectral fits. The $\Delta_{\rm AIC}$ column gives the AIC relative to the best fitting model, which is 2SBPL unless $\Delta_{\rm AIC} = 0$ (though note that even in these cases, $\Delta_{\rm AIC} < 4$ for the 2SBPL model in all epochs except 36s, where $\Delta_{\rm AIC} = 4.3$). $\Delta_{\rm AIC} > 4$ means that the Band+BB model is disfavoured, while $\Delta_{\rm AIC} > 10$ indicates no support for the model at all. Data are presented as mean values ± one standard deviation.}
\label{tab:bandbbfits}
\end{table*}

\begin{table*}
\small
\centering 
\begin{tabular}{@{}ccccccc} 
\hline\hline
Model & $\beta$ & $N_{\rm H}$ & $E(B-V)$ & $E_{b}$ & $\chi ^2$ (dof) & Null Hypothesis\\ 
 & & ($10^{20} {\rm cm^{-2}}$) & (mag) & (keV) & & Probability\\ 
\hline
 MW/POW  & $ 1.22 ^{+ 0.03 }_{ -0.03 } $ & $<3.45$ & $ 0.03 ^{+ 0.03 }_{ -0.03 } $ & $ - $ & 53 (32)  &  0.012 \\
 LMC/POW  & $ 1.22 ^{+ 0.03 }_{ -0.03 } $ & $<3.45$ & $ 0.03 ^{+ 0.03 }_{ -0.03 } $ & $ - $ & 53 (32)  &  0.011 \\
 SMC/POW  & $ 1.21 ^{+ 0.02 }_{ -0.02 } $ & $<3.35$ & $ 0.02 ^{+ 0.03 }_{ -0.02 } $ & $ - $ & 53 (32)  &  0.01 \\
 MW/BKP  & $ 1.48 ^{+ 0.11 }_{ -0.07 } $ & $ 0.74 ^{+ 4.00 }_{ -0.74 } $ & $ 0.06 ^{+ 0.04 }_{ -0.04 } $ & $ 0.02 ^{+0.02}_{-0.01} $ & 34 (31)  &  0.341 \\
 LMC/BKP  & $ 1.48 ^{+ 0.11 }_{ -0.07 } $ & $ 0.74 ^{+ 4.00 }_{ -0.74 } $ & $ 0.06 ^{+ 0.03 }_{ -0.03 } $ & $ 0.02 ^{+0.02}_{-0.01} $ & 33 (31)  &  0.355 \\
 SMC/BKP  & $ 1.48 ^{+ 0.11 }_{ -0.07 } $ & $ 0.74 ^{+ 4.00 }_{ -0.74 } $ & $ 0.05 ^{+ 0.03 }_{ -0.03 } $ & $ 0.03 ^{+0.02}_{-0.01} $ & 33 (31)  &  0.357 \\
\hline\hline
\end{tabular}
\caption{Results from simultaneous UV/optical and X-ray spectral fits for the SMC, LMC and MW dust-extinction law models, for both a power-law (POW) and a broken power-law (BKP) continuum to the $5.5~{\rm ks}$ spectral energy distribution. The lower energy photon index of broken-power-law model is fixed to be $\alpha=2/3$, the $\beta$ listed below is for the higher energy spectral index. The columns are: the model, photon index $\beta$, the host galaxy equivalent column density (upper limits are given at $3\sigma$), host $E(B-V)$, the break energy for the broken power-law $E_b$, the $\chi ^2$ and degree of freedom (dof) of the fit, and finally the null hypothesis probability. We take the fit with the highest value of $E(B-V)$ (MW/BKP) for our $3\sigma$ limit in the main text. Data are presented as mean values ± one standard deviation.
\label{tab:sedfits}}
\end{table*}

\clearpage
